\documentclass[12pt]{article}
\usepackage[utf8]{inputenc}
\usepackage{amsmath, amssymb, amsthm, mathrsfs, mathtools,bm}
\usepackage{slashed}
\usepackage[colorlinks=true, linkcolor=black, citecolor=black, urlcolor=blue]{hyperref}
\usepackage{tikz}
\usetikzlibrary{cd}
\usepackage{graphicx}
\usetikzlibrary{decorations.markings, decorations.pathreplacing, calc, arrows.meta, positioning}
\usetikzlibrary{shapes.geometric}
\usepackage{titlesec}

\theoremstyle{plain}

\theoremstyle{definition}

\theoremstyle{remark}
\newtheorem{remark}{Remark}

\usepackage[left=2.5cm,right=2.5cm,top=2.5cm,bottom=3cm]{geometry}
\fontdimen2\font=1.2\fontdimen2{\jot}{5pt}

\newcommand{\field}[1]{\ensuremath{\mathbb{#1}}}
\newcommand{\CC}{\field{C}}

\newcommand{\RR}{\field{R}}

\newcommand{\ZZ}{\field{Z}}
\newcommand{\beq}{\begin{equation}\begin{aligned}}
\newcommand{\eeq}{\end{aligned}\end{equation}}

\newcommand{\be}{\beta}
\newcommand{\del}{\delta}

\newcommand{\ov}{\over}

\newcommand{\curly}[1]{\mathscr{#1}}

\newcommand{\cD}{\curly{D}}

\newcommand{\lam}{\lambda}

\numberwithin{equation}{section}

\titleformat{\section}{\large\bfseries}{\thesection.}{4pt}{}
\titlespacing{\section}{0pt}{20pt}{6pt}

\titleformat{\subsection}{\normalfont\bfseries}{\thesubsection.}{4pt}{}
\titlespacing{\subsection}{0pt}{15pt}{6pt}

\titleformat{\subsubsection}{\normalfont\itshape}{\thesubsubsection.}{4pt}{}
\titlespacing{\subsubsection}{0pt}{15pt}{6pt}

\titleformat{\paragraph}{\normalfont\itshape}{\theparagraph.}{4pt}{}
\titlespacing{\paragraph}{0pt}{15pt}{6pt}

\DeclareFontShape{OT1}{cmr}{mx}{n}%
{<->cmr10}{}
\newcommand{\mytitlefont}{\fontseries{mx}\selectfont}
\DeclareMathAlphabet{\titlemath}{OT1}{cmr}{mx}{n}

\begin{document}

\begin{titlepage}
%\begin{flushright} \small
%UUITP-17/19
 %\end{flushright}

\begin{center}
			
~\\[0.9cm]
			
{\fontsize{27pt}{0pt} \mytitlefont  Worldline Localization}

~\\[1cm]

{\fontsize{15pt}{20pt}\selectfont 
Changha Choi$^{a}$\footnote{\href{mailto:cchoi@perimeterinstitute.ca}{\tt cchoi@perimeterinstitute.ca}},
\;
Leon A.\ Takhtajan$^{b,c}$\footnote{\href{mailto:leontak@math.stonybrook.edu}{\tt leontak@math.stonybrook.edu}}
}

~\\[0.5cm]

{\it 
$^{a}$Perimeter Institute for Theoretical Physics,\\
Waterloo, Ontario, N2L 2Y5, Canada
}\\[0.25cm]

{\it 
$^{b}$Department of Mathematics, Stony Brook University,\\
Stony Brook, NY 11794, USA\\
$^{{c}}$Euler International Mathematical Institute,\\
Pesochnaya Nab. 10, Saint Petersburg 197022 Russia
}

\end{center}

\vskip0.5cm
			
\noindent 

\makeatletter
\renewenvironment{abstract}{%
    \if@twocolumn
      \section*{\abstractname}%
    \else %% <- here I've removed \small
      \begin{center}%
        {\bfseries \normalsize\abstractname\vspace{\z@}}%  %% <- here I've added \Large
      \end{center}%
      \quotation
    \fi}
    {\if@twocolumn\else\endquotation\fi}
\makeatother

\begin{abstract}
We show that two elementary worldline path integrals—the thermal
partition function of the harmonic oscillator and the one–loop effective
action of scalar QED in a constant field strength—exhibit a natural form
of supersymmetric localization.  The mechanism relies on hidden
fermionic symmetries of the worldline BRST formulation, rather than on
standard BRST structure or physical supersymmetry.  These symmetries
localize the target–space trajectory.  For the harmonic oscillator this
yields an alternative localization derivation of the Jacobi–Poisson
formula.  Moreover, after the trajectory is localized, the remaining
proper–time integral exhibits an emergent zero–dimensional supersymmetry
generated by modular invariance, allowing the modulus \(T\) itself to be
localized.  For scalar QED the same structure provides a controlled
computation of both the real and imaginary parts of the Euler–Heisenberg
effective action.  In particular, the imaginary part arises from a
moduli space of circular worldline instantons, offering a localization
perspective on the semiclassical exactness of the Schwinger
pair-production result observed by Affleck–Alvarez–Manton.
\end{abstract}
\vfill
\end{titlepage}

\tableofcontents

\section{Introduction}

Worldline path integrals \cite{Feynman:1950ir,Schwinger:1951nm} provide a
powerful first–quantized representation of quantum field theory (see
\cite{Schubert:2001he} for a review).  
The simplest example is the one–loop effective action obtained from the
Fock–Schwinger proper–time representation
\cite{Fock:1937dy,Schwinger:1951nm}, whose evaluation already captures a
remarkably rich set of physical phenomena.  
In QED, for instance, the one–loop worldline formalism reproduces
nonlinear quantum–vacuum effects such as light–light scattering, as well
as the nonperturbative Schwinger mechanism for pair production in strong
electric fields.

The purpose of this paper is to reveal that, for two of the simplest
examples of worldline path integrals — the thermal partition function of
the harmonic oscillator and the one–loop effective action of scalar QED
in a constant electric field — the standard computations admit a natural
interpretation in terms of supersymmetric localization.  
We demonstrate that the worldline formalism contains hidden fermionic
symmetries that allow these integrals to be evaluated by localization
techniques.

These fermionic symmetries are not related to physical supersymmetry.
Rather, they emerge when the worldline path integral is written in its
reparameterization-invariant form and quantized through the BRST/BV
formalism.  While the standard BRST symmetry encodes the gauge
invariance of this formulation, we find that the construction contains
additional hidden fermionic symmetries.  These hidden symmetries provide
a natural mechanism for the localization of the worldline path integral.

The worldline path integral contains two types of variables: the proper
time \(T\), which plays the role of the worldline modulus, and the
target–space trajectory \(x^\mu(\tau)\).  
In both examples studied in this paper, we show that the integration
over \(x^\mu\) is localized by the hidden fermionic symmetries of the
worldline formalism.  

For the harmonic oscillator, this provides an alternative localization
derivation of the Jacobi–Poisson formula \cite{Choi:2021yuz}.  
Moreover, we find that once the \(x\)-integral is localized, the
remaining integration over the modulus \(T\) can be performed by a
zero–dimensional localization argument that exploits the modular
symmetry of the partition function.

For scalar QED in a constant electric field background, the hidden fermionic
symmetries localize the \(x^\mu\)–path integral onto the Lorentzian
equations of motion.  
For generic values of \(T\) the localization critical point is a trivial
constant map, which reproduces the real part of the
Euler–Heisenberg integrand.  
At a discrete set of resonant values of \(T\), however, the equations of
motion admit a nontrivial moduli space of worldline instantons --- the
cyclotron trajectories — which contribute an imaginary part to the
effective action and thus encode the vacuum decay rate.

The localization derivation of the scalar QED effective action provides a
controlled computation of both $\mathrm{Re}\,\Gamma$ and
$\mathrm{Im}\,\Gamma$, and in doing so clarifies the notable observation
of Affleck–Alvarez–Manton (AAM) \cite{Affleck:1981bma}.  
AAM discovered that the imaginary part of the effective action — which
encodes the Schwinger pair-production rate — is exactly reproduced by a
semiclassical stationary-point analysis in both the worldline coordinates
$x^\mu$ and the proper time \(T\).  
In contrast to the AAM setup, our localization locus contains a
nontrivial moduli space, and this structural difference plays an
important role in understanding the emergence of the imaginary
contribution.

The organization of the paper is as follows.  
In Section~\ref{sec:HOsemi}, we discuss the exact semiclassicality of the harmonic oscillator.  
In Section~\ref{sec:BRST}, we present the BRST/BV formulation of the spinless particle.  
In Section~\ref{sec:HOloc}, we introduce the hidden fermionic symmetries and localize the harmonic oscillator worldline path integral.  
In Section~\ref{sec: Schwinger}, we set up the worldline representation of scalar QED in a constant field strength.  
In Section~\ref{sec:Schwingerloc}, we apply the corresponding fermionic symmetries to scalar QED and derive its one–loop effective action via worldline localization.  
In Section~\ref{sec:0d-LG-toy}, we present a zero–dimensional toy model that illustrates the Euclidean off-shell supersymmetry and localization structure in 0d Morse theory.

\section*{Acknowledgments}  The first author (C.C.) thanks Kevin Costello, Jaume Gomis, Shota Komatsu, and Sungjay Lee for helpful discussions. 
C.C. further thanks CERN for its hospitality during the period in which part of this work was completed. 
The research of C.C. was supported by the Perimeter Institute for Theoretical Physics. 
Research at Perimeter Institute is supported in part by the Government of Canada through the Department of Innovation, Science and Economic Development and by the Province of Ontario through the Ministry of Colleges and Universities.

\section{Exact Semiclassicality of the Simplest Quantum System}
\label{sec:HOsemi}

The one-dimensional harmonic oscillator is considered as the most elementary exactly solvable system in quantum mechanics.  
A single one-loop (Gaussian) stationary phase evaluation reproduces the full quantum result with no higher corrections.  
Yet, while reexamining its partition function using the Fock-Schwinger proper time formalism, we find that an unexpected structure emerges:  
the system is \emph{exactly semiclassical}  both in proper time and in worldline variables.

\subsection{The Setup}
We start from the Euclidean Lagrangian for the harmonic oscillator,
\begin{equation}
\mathcal L = \frac{1}{2}\,\phi\,(\Box+ m^2)\,\phi, 
\qquad \Box=-\partial_x^2.
\end{equation}
Its partition function is given by the textbook formula:
\begin{equation}\label{Z-beta}
Z(\beta)=\sum_{n=0}^{\infty}e^{-\beta E_{n}}
   = \frac{1}{2\sinh(\tfrac{\beta m}{2})},
\end{equation}
where $\beta>0$ is the inverse temperature, and $E_{n}=m(n+\tfrac{1}{2})$ are the energy levels.

On the other hand, the thermal trace and the free energy can be computed using the Euclidean path integral, which is one-loop exact:
\begin{align}
Z(\beta)
   &=\int \mathscr D\phi\,
     e^{-\frac{1}{2}\int_0^\beta \phi (-\partial_x^2+m^2)\phi\,dx}
     = \frac{1}{\sqrt{\det(\Box+m^2)}},\\[4pt]
W(\beta)&=-\log Z(\beta)
   = \frac{1}{2}\log\det(\Box+m^2).
\end{align}
Here $\det(\Box+m^2)$ is the zeta function regularized determinant of the differential operator $\Box+m^{2}$ on the thermal circe $S^{1}_{\beta}=\RR/\be\ZZ$.

It is instructive to express the free energy in the worldline formalism using the Fock-Schwinger's proper time method:
\begin{equation}
W(\beta)
   \simeq -\frac{1}{2}\!\int_0^\infty\!\frac{dT}{T}
      \!\int_{\text{Map}(S^{1}_{T},S^{1}_{\be})}\!\!\mathscr D x\,
      e^{-\int_0^T \frac{1}{2}(\dot x^2+m^2)\,d\tau},
\label{eq:SchwingerPath}
\end{equation}
where domain of integration consists of maps $x(\tau)$ with the property that $e^{ix(\tau)}$ maps $S^{1}_{T}$ to $S^{1}_{\be}$. The symbol $\simeq$ indicates that UV divergence at the lower limit $T=0$ should be renormalized, as we will discuss it momentarily.

After rescaling $\tau\!\to\!T\tau$ we obtain
\begin{equation}
W(\beta)
   \simeq - \frac{1}{2}\!\int_0^\infty\!\frac{dT}{T}
      \!\int_{\text{Map}(S^{1},S^{1}_{\be})}\!\!\mathscr D x\,
      e^{-\int_0^1 \frac{1}{2}\!\left(\frac{\dot x^2}{T}+T m^2\right)\!d\tau},
\label{eq:RescaledPath}
\end{equation}
where $S^{1}=\RR/\ZZ$, so $x(\tau)$ is a worldline variable defined on the unit interval,
while $T$ is a global modulus integrated with the measure $dT/T$.
Its simultaneous appearance in the kinetic term (as $1/T$) and in the
mass term (as $T$) will be central to what follows.

\subsection{Classical Solutions and Gaussian Evaluation}
The critical points of the action in \eqref{eq:RescaledPath} satisfy
\begin{equation}\label{cr-points}
\ddot x(\tau)=0,
\qquad
\int_0^1\!\dot x^2 d\tau = m^2T^2,
\end{equation}
so corresponding solutions are labeled by an integer $n$, the winding number of a map $e^{ix(\tau)}$,
\begin{equation}
x_n(\tau)=x_0+n\beta\tau,
\qquad
T_n=\frac{|n|\beta}{m}.
\end{equation}
Thus both the worldline coordinate $x(\tau)$ and the proper-time modulus $T$
localize to discrete classical configurations labeled by $\ZZ$.

Corresponding worldline path integral in  \eqref{eq:RescaledPath} is Gaussian, and as it was shown in \cite{Choi:2021yuz}, it localizes to the critical points $x_{n}(\tau)$. Thus we obtain
\begin{equation}
W(\beta)
   \simeq - \frac{\beta}{2(2\pi)^{1/2}}
     \!\int_0^\infty\!\!\frac{dT}{T^{3/2}}
     \sum_{n\in\mathbb Z}
       e^{-\frac{1}{2}(m^2T+n^2\beta^2/T)}.
\label{eq:IntegralForm}
\end{equation}
For $n\neq0$ each integral in the sum is convergent, and the exponent
$f_n(T)=\tfrac{1}{2}(m^2T+n^2\beta^2/T)$
has a unique critical point 
\begin{equation}
T_n=\frac{|n|\beta}{m},
\qquad
f_n(T_n)=|n|\beta m,
\qquad
f_n''(T_n)=\frac{m^3}{|n|\beta}.
\end{equation}
Corresponding stationary point approximation is \emph{exact} and gives
\begin{equation}
\int_0^\infty\!\frac{dT}{T^{3/2}}e^{-f_n(T)}
   = \frac{(2\pi)^{1/2}}{|n|\beta}\,
   e^{-|n|\beta m}.
\end{equation}
Substituting into \eqref{eq:IntegralForm}, we obtain
\begin{equation}
W(\be)|_{n\neq 0}
   = -\frac{1}{2}\!
   \sum_{n\neq0}\!\frac{e^{-|n|\beta m}}{|n|}
   = \log\! \left(1-e^{-\be m}\right).
   \end{equation}

\subsection{UV Regularization and Renormalization}
The $n=0$ sector is UV divergent and must be renormalized.  
We define the regularized free energy $W(\beta;\Lambda)$ by introducing a hard proper time cutoff in \eqref{eq:RescaledPath}:
\begin{equation}
W(\beta;\Lambda)
   =- \frac{1}{2}\!\int_{\Lambda^{-2}}^\infty\!\frac{dT}{T}
      \!\int_{\text{Map}(S^{1},S^{1}_{\be})}\!\!\mathscr D x\,
      e^{-\int_0^1 \frac{1}{2}\!\left(\frac{\dot x^2}{T}+T m^2\right)\!d\tau}.
\end{equation}
Then the $n=0$ contribution becomes
\begin{equation}
W(\beta;\Lambda)|_{n=0}
   =-\frac{\beta}{2(2\pi)^{1/2}}
     \int_{\Lambda^{-2}}^\infty\!\frac{dT}{T^{3/2}}\,e^{-\tfrac{1}{2}m^2 T}
   =-\frac{\beta\Lambda}{\sqrt{2\pi}}
     +\frac{\beta m}{2}
     +O(1/\Lambda).
\label{eq:n0HOreg}
\end{equation}

To remove the UV divergence, we introduce a local counterterm
\[
\Delta W(\Lambda)
   = C(\Lambda)\!\int_{S^1_\beta}\!\sqrt{g}\,dx
   = C(\Lambda)\,\beta,
\]
whose coefficient $C(\Lambda)$ is chosen to cancel the linear divergence
in \eqref{eq:n0HOreg}.  
This condition implies
\begin{equation}
C(\Lambda) - \frac{\Lambda}{\sqrt{2\pi}} = O(\Lambda^{0}).
\end{equation}

There remains a finite scheme ambiguity corresponding to the constant term part
$O(\Lambda^0)$ of $C(\Lambda)$, which can be fixed by requiring
that the vacuum energy of the harmonic oscillator to be $E_0 = m/2$.
This condition uniquely determines the counterterm coefficient as
\begin{equation}
C(\Lambda) = \frac{\Lambda}{\sqrt{2\pi}}.
\end{equation}

With this physical renormalization scheme, we obtain
\begin{equation}
W_{\text{ren}}(\beta)
   = \lim_{\Lambda\to\infty}(W(\beta;\Lambda)+\Delta W(\Lambda))
   = \log\left[2\sinh(\tfrac{\beta m}{2})\right],
\end{equation}
which precisely reproduces the free energy obtained from the exact spectrum.

\vspace{1em}
Thus the harmonic oscillator is not only exactly solvable but is
\emph{semiclassically exact} with respect to both the worldline
coordinates and the proper time modulus.  
No higher corrections appear: the full partition function is obtained
solely from the classical loops and the integration over their proper time
modulus.

The localization of the dynamical variables \(x(\tau)\) may be viewed as
a worldline realization of the non–supersymmetric localization
mechanism developed in \cite{Choi:2021yuz}.  
However, in the present setting a new phenomenon occurs: the proper time
variable \(T\) also localizes.  
This naturally raises the question of whether there exists a
zero–dimensional fermionic symmetry — emerging from modular invariance — that
localizes the modulus \(T\) itself.

%%%%%%%

\section{BRST Quantization of the Spinless Particle}
\label{sec:BRST}

The Fock--Schwinger proper time representation of the bosonic harmonic
oscillator can be reformulated as a one–dimensional gauge theory by
introducing an einbein \(e(\tau)\) on the worldline, with the gauge
symmetry given by worldline reparametrizations.
In this section we quantize this gauge theory in the BRST formalism,
which naturally introduces the corresponding fermionic degrees of
freedom (ghost and antighost fields).

We will identify the resulting fermionic symmetries and explain how they
underlie the semiclassical exactness observed in the previous section.

\subsection{Gauge Symmetry and Abelianization}
We start from the standard einbein action for a relativistic spinless massive particle, 
\begin{equation}
S_0[x,e]
   = \frac{1}{2}\!\int_0^1\!d\tau\,
      \Big(e^{-1}\dot x^2 + e\,m^2\Big),
\label{eq:S0}
\end{equation}
where $x^\mu(\tau)$ is the worldline embedding and $e(\tau)>0$ is the einbein.  
The action is invariant under an arbitrary reparametrization of the worldline parameter,
\[
\tau \;\longrightarrow\; \tau'(\tau), 
\qquad
x^\mu(\tau) \;\to\; x'^{\mu}(\tau') = x^\mu(\tau), 
\qquad
e(\tau)\;\to\; e'(\tau') = e(\tau)\,\frac{d\tau}{d\tau'},
\]
which corresponds to a worldline diffeomorphism.  
Linearizing this transformation gives the infinitesimal form
\begin{equation}
\delta_\epsilon x^\mu = \epsilon\,\dot x^\mu,
\qquad
\delta_\epsilon e = \partial_\tau(\epsilon e),
\label{eq:reparam-nonabelian}
\end{equation}
with an arbitrary gauge parameter $\epsilon(\tau)$.
The commutator of two such transformations closes with a field-dependent parameter,
\(
[\delta_{\epsilon_1},\delta_{\epsilon_2}] = \delta_{\epsilon_1\dot\epsilon_2-\epsilon_2\dot\epsilon_1},
\)
so in this form the gauge Lie algebra is not abelian.

Following \cite{Gomis:1994he}, it is convenient to \emph{abelianize} the symmetry by redefining the gauge parameter to be 
\begin{equation}
\tilde\epsilon \equiv e\,\epsilon.
\end{equation}
In terms of $\tilde\epsilon(\tau)$ the symmetry \eqref{eq:reparam-nonabelian} becomes
\begin{equation}
\delta_{\tilde\epsilon} x^\mu = e^{-1}\tilde\epsilon\,\dot x^\mu,
\qquad
\delta_{\tilde\epsilon} e = \partial_\tau \tilde\epsilon,
\label{eq:reparam-abelian}
\end{equation}
and now the transformations commute:
\(
[\delta_{\tilde\epsilon_1},\delta_{\tilde\epsilon_2}]=0.
\)
From now on we treat \eqref{eq:reparam-abelian} as our gauge symmetry.  
This allows us to use the usual BRST quantization for abelian gauge theories.

\subsection{Gauge Fixing and BRST Quantization}

To BRST–quantize the worldline theory we introduce, for the abelianized
reparametrization symmetry, a ghost $C(\tau)$, an antighost
$\bar C(\tau)$, and an auxiliary Nakanishi--Lautrup field $b(\tau)$.
The BRST operator acts as
\begin{equation}
\boxed{
\begin{aligned}
s\,x^\mu &= e^{-1} C\,\dot x^\mu,\\[2pt]
s\,e &= \dot C,\\[2pt]
s\,C &= 0,\\[2pt]
s\,\bar C &= b,\\[2pt]
s\,b &= 0 .
\end{aligned}}
\label{eq:BRST-basic-rules}
\end{equation}
These transformations satisfy $s^{2}=0$ off-shell.

\medskip

We fix the reparametrization symmetry using the \emph{derivative gauge}
\begin{equation}
\dot e(\tau)=0 ,
\label{eq:derivative-gauge}
\end{equation}
which removes all $\tau$–dependent fluctuations of the einbein and leaves
its constant mode unfixed.

The gauge–fixing fermion is chosen as
\begin{equation}
\Psi
= -\!\int_0^1 d\tau\, \bar C\,\dot e .
\label{eq:gauge-fixing-fermion}
\end{equation}

Acting with the BRST operator yields
\begin{equation}
s\Psi
=
\int_0^1 d\tau\,
\Big[
-\,b\,\dot e
 - \dot{\bar C}\,\dot C
\Big].
\end{equation}

Thus the full gauge–fixed action becomes
\begin{equation}
S_{\mathrm{BRST}}
=
S_0[x,e]
+
s\Psi
=
\int_0^1 d\tau\,
\Big[
\frac12\big(e^{-1}\dot x^{2} + e\,m^{2}\big)
-\,b\,\dot e
 - \dot{\bar C}\,\dot C
\Big].
\label{eq:Sgf-1}
\end{equation}
In Euclidean signature the $b$–integration is taken along the imaginary
axis, so that integrating out $b$ imposes $\delta[\dot e]$.  The remaining
ghost term $-\,\dot{\bar C}\,\dot C$ is the correct Faddeev–Popov
determinant for the derivative gauge.

%\medskip
%\noindent\textbf{Zero-mode multiplets and extended BRST.}
\subsection{Residual Gauge Fixing}

The gauge condition $\dot e=0$ removes all $\tau$–dependent fluctuations
of the einbein but leaves unfixed the constant (zero–mode) parts of the
ghosts.  The ghost system still has
two fermionic shift symmetries,
\begin{equation}
C(\tau)\to C(\tau)+\epsilon_{C},\qquad 
\bar C(\tau)\to \bar C(\tau)+\epsilon_{\bar C},
\label{eq:resi}
\end{equation}
with constant Grassmann parameters.  
These residual symmetries generate ghost and antighost zero modes,
causing the naive BRST path integral to vanish.  
They must therefore be gauge–fixed separately.

This is naturally handled by the Batalin–Vilkovisky (BV) formalism
\cite{Batalin:1981jr,Batalin:1983ggl,Henneaux:1992ig,Gomis:1994he,costello2011renormalization}.  
The residual shift of $C$ requires a bosonic ghost-of-ghost and a
fermionic auxiliary field, while the residual shift of $\bar C$ requires
a bosonic extraghost and a fermionic auxiliary field.

However, for our purposes it is enough to take a simple shortcut (which also arises from the BV prescription) that directly fixes
the constant shift symmetries by inserting the corresponding fermionic
operators.  
This leads to the gauge–fixed action
\begin{equation}
S_{\mathrm{BRST}}
=
\int_0^1 d\tau\left[
\frac12\big(e^{-1}\dot x^2 + e\,m^2\big)
- b\,\dot e
- \dot{\bar C}\,\dot C
\right]
\;+\;
C_0\,\eta+\bar C_0\,\xi ,
\label{eq:Sgf-full-smooth}
\end{equation}
where
\[
C_0=\int_0^1\! d\tau\, C(\tau),\qquad
\bar C_0=\int_0^1\! d\tau\, \bar C(\tau),
\]
and $\eta,\xi$ are constant (0-dimensional) fermionic auxiliary fields.
Integrating out $\eta$ and $\xi$ inserts the factor $C_0\bar C_0$, which
precisely fixes the residual fermionic gauge symmetry
\eqref{eq:resi}.

Integrating out $b$ imposes $\dot e=0$, reducing $e(\tau)$ to its
constant mode $T$.  
The resulting reduced action is
\begin{equation}
S_{\mathrm{BRST}}^{\mathrm{red}}
=
\int_0^1 d\tau\left[
\frac12\big(T^{-1}\dot x^2 + T\,m^2\big)
- \dot{\bar C}\,\dot C
\right]
\;+\;
C_0\,\eta+\bar C_0\,\xi,
 \label{eq:Sgf-reduced}
\end{equation}
which is the form used in the localization analysis.

\subsection{BRST Gauge–Fixed Path Integral}

In the derivative gauge $\dot e = 0$, the Faddeev--Popov operator is
$T\,\partial_\tau^{2}$.  
After dividing by the volume of the residual global reparametrization
(the zero mode of the gauge parameter), the functional measure for the
einbein reduces to
\[
\frac{\cD e}{\mathrm{Vol}(\mathrm{Diff}(S^{1}))}
\;\longrightarrow\;
\frac{dT}{T}.
\]

After integrating out $b$ and imposing $\dot e = 0$, the gauge-fixed BV
path integral becomes
\begin{equation}
\int_0^\infty\!\frac{dT}{T}
\int\!\mathscr{D}x\,\mathscr{D}C\,\mathscr{D}\bar C\, d\xi\, d\eta\;
e^{-S_{\mathrm{BRST}}^{\mathrm{red}}[x,C,\bar C,\xi,\eta]}.
\end{equation}
Thus we see that up to an overall irrelevant normalization, the BRST/BV
quantization of the spinless worldline particle reproduces the harmonic–oscillator
free energy in the Fock–Schwinger proper–time representation,
Eq.~\eqref{eq:RescaledPath}.

\section{Worldline localization for the Harmonic Oscillator} \label{sec:HOloc}
In this section we identify two nontrivially hidden fermionic symmetries in the BRST/BV action \eqref{eq:Sgf-reduced}, distinct from the standard BRST symmetry inherited from the gauge invariance.  
The key observation is that the fields in the BRST/BV action naturally organize into two layers of fermionic structure: a one-dimensional supersymmetry acting on the oscillatory (nonzero-mode) sector, and a zero-dimensional supersymmetry acting on the constant term (modulus) sector.  
The crucial step is to treat the oscillatory and constant term components of the ghost and antighost fields as independent degrees of freedom --- equivalently, to regard $\dot C$ and $C_0$ (and likewise $\dot{\bar C}$ and $\bar C_0$) as independent variables in the functional integral.

According to this two level structure, we implement the localization in two steps. 
First we perform the localization at the level of the target space variable $x$ (for simplicity we are assuming the one-dimensional target space), 
and then subsequently localize the corresponding modulus $T$. 
This two-step process provides a localized reformulation of the semiclassical analysis 
presented in Section~\ref{sec:HOsemi}.
\subsection{Localization of the Target Space Modes}  \label{sec: 1st layer}

The gauge-fixed action \eqref{eq:Sgf-reduced} admits two one-dimensional fermionic symmetries acting on the fields $(x, \dot C, \dot{\bar C})$, given by\footnote{%
These symmetries rely crucially on the higher–derivative ghost term
$\int d\tau\,\dot{\bar C}\,\dot C$.
They have no analogue in a theory whose fermions appear with the usual
first–order kinetic term $\int d\tau\,\bar C\,\dot C$, such as the
$\mathcal N=1$ worldline supersymmetry considered in~\cite{Choi:2021yuz,Choi:2023pjn,Choi:2025iqp} (and see \cite{Murthy:2025ioh} for 2d analogue),
where localization arises through a different mechanism.
}
\begin{equation}
\begin{aligned}
&\delta_1 x = T^{\frac{3}{2}}\,\dot{\bar C}, 
&&\delta_1 \dot C = T^{\frac{1}{2}}\,\ddot x,
&&\delta_1 \dot{\bar C} = 0,  \\[4pt]
&\bar\delta_1 x = T^{-\frac{1}{2}}\,\dot{\bar C}, 
&&\bar\delta_1 \dot C = 0 ,
&&\bar\delta_1 \dot{\bar C} = -\,T^{-\frac{3}{2}}\,\ddot x.
\end{aligned}
\label{eq:susy1os}
\end{equation}
In the following we focus on $\delta_1$; the analysis using $\bar\delta_1$ can be performed in parallel.
It is important to note that $\delta_1$ acts only on the one-dimensional (oscillatory) sector spanned by $(x, \dot C, \dot{\bar C})$, 
while leaving the zero-dimensional variables $(C_0,\bar C_0,T,\xi,\rho,\eta)$ inert. 
In particular, $\delta_{1}C_{0}=\delta_{1}\bar{C}_{0}=0$ for the constant temrs $C_0$ and $\bar C_0$ associated with the reparametrization ghosts. Also, $\delta_{1}x_{0}=0$ for the constant term of $x(\tau)$.  

The powers of $T$ in \eqref{eq:susy1os} are fixed by the dimensional analysis.  
Taking $[m]=+1$ and $[\tau]=0$, we get the following formulas for the dimensions of other variables,
\begin{equation}
[x]=-1,\qquad [e]=-2,\qquad [T]=-2, \qquad
[C]=-2, \qquad
[\bar C]=+2.
\label{eq:dimensions}
\end{equation}
They easily follow from the definition $T=\!\int_0^1 e(\tau)\,d\tau$, the scaling of the reparametrization ghost, and agree with the kinetic term $\int_{0}^{1} \dot{\bar C}\,\dot C d\tau$ being dimensionless.

\vspace{0.5em}
However, the symmetry \eqref{eq:susy1os} does not close of-shell.  To close the supersymmetry algebra,
we introduce an auxiliary field $G$ by inserting it into the path integral through a trivial Gaussian measure
\[
1=\int \mathcal \cD G \;
e^{-\frac{1}{2}\int_0^1G^2\,d\tau},
\]
which assigns $[G]=0$. Correspondingly, we extend the BV action $ S_{\mathrm{BRST}}^{\mathrm{red}}$
by adding the auxiliary Gaussian term,
\begin{equation}
S_{\mathrm{off}}
   = S_{\mathrm{BRST}}^{\mathrm{red}}
   + \frac{1}{2}\!\int_0^1\! d\tau\, G^2 .
\label{eq:SBVex}
\end{equation}

The extended off-shell supersymmetry generated by~$\delta_1$ then reads\footnote{
Throughout this paper we use the same symbol~$\delta$ 
for a supersymmetry transformation and its corresponding off-shell extension; 
the distinction will be clear from context.
}:
\begin{equation}
\boxed{
\begin{aligned}
\delta_1 x &= T^{\frac{3}{2}}\,\dot{\bar C}, 
&\qquad \delta_1 \dot C &= T^{\frac{1}{2}}\,\ddot x +i T\,\dot G, \\[6pt]
\delta_1 \dot{\bar C} &= 0, 
&\qquad \delta_1 G &= i\,T\,\ddot{\bar C}.
\end{aligned}
}
\label{eq:delta1_offshell}
\end{equation}
This transformation satisfies $\delta_1^2 = 0$ and leaves the action  $S_{\mathrm{off}}$ invariant,
\[
\delta_1 S_{\mathrm{off}}= 0.
\]

Following the philosophy in  \cite{Choi:2021yuz}, we introduce a $\delta_1$-exact higher derivative deformation by
\begin{equation}
V_1
= T^{-\frac{3}{2}}\!\int_0^1\!d\tau\,\ddot x\,\dot C,
\qquad
S^{\text{loc}}_1(\lambda_1)
=\lambda_1\,\delta_1 V_1 .
\end{equation}
A direct application of $\delta_1$ yields
\begin{equation}
\lambda_1\,\delta_1 V_1
= \lambda_1 T^{-1}\!\int_0^1\!d\tau\,(\ddot x)^2
+ i\,\lambda_1 T^{-\frac{1}{2}}\!\int_0^1\!d\tau\,\ddot x\,\dot G
+ \lambda_1\!\int_0^1\!d\tau\,\dddot{\bar C}\,\dot C .
\label{eq:delta1V1-nosign}
\end{equation}
Integrating out $G$ completes the square and contributes
\(\tfrac{\lambda_1^{2}}{2}\,T^{-1}\!\int(\dddot x)^2\),
leading to following the effective deformation
\begin{equation}
\boxed{
S^{\text{loc}}_1(\lambda_1)
= \lambda_1 T^{-1}\!\int_0^1\!d\tau\,(\ddot x)^2
+ \frac{\lambda_1^{2}}{2}\,T^{-1}\!\int_0^1\!d\tau\,(\dddot x)^2
+ \lambda_1\!\int_0^1\!d\tau\,\dddot{\bar C}\,\dot C \,.
}
\label{eq:Sdef-final-nosign}
\end{equation}
To apply the localization by sending $\lambda_{1}\to\infty$, both the linear and quadratic $\lambda_1$-dependent
bosonic terms must have positive real part. This requires
\[
\text{Re}\,\lambda_1>0,
\qquad
\text{Re}\,\lambda_1^2>0,
\]
which together imply $\lambda_{1}$ is in the wedge
\[
\boxed{-\,\frac{\pi}{4} < \arg\lambda_1 < \frac{\pi}{4}.}
\]
In this region of the complex plane, the deformation
$S^{\mathrm{loc}}_1(\lambda_1)$ in the limit $\lambda_{1}\to\infty$ exponentially suppresses all
configurations except classical trajectories $\ddot x= 0$.  
Thus the path integral therefore 
localizes to the classical solutions
\beq \label{eq:xsaddle}
x_n(\tau)=x_0+n\,\beta\,\tau,
\qquad n\in\mathbb{Z},
\eeq
parametrized by the winding numbers of the target space variable.

Now the computation of the functional determinant proceeds exactly as in
\cite{Choi:2021yuz} (despite the difference between the
$\mathcal{N}=\tfrac12$ and $\mathcal{N}=1$ fermionic symmetries).  
The deformation~\eqref{eq:Sdef-final-nosign} localizes the field $x$ to
the classical trajectories~\eqref{eq:xsaddle} and completely integrates
out the oscillatory ghost modes $C_{\mathrm{osc}}$ and
$\bar C_{\mathrm{osc}}$.  
We therefore obtain
\begin{equation}
\label{mail-target-space}
\begin{gathered}
\log Z(\beta)
= \frac{1}{2}\!\int_0^\infty\!\frac{dT}{T}
   \!\int\!\mathscr D x\,\mathscr D C\,\mathscr D\bar C\, d\xi\, d\eta\;
   e^{-S_{\mathrm{off}}[x,C,\bar C,\xi,\eta]}
\\[4pt]
=\frac{\beta}{2(2\pi)^{1/2}}
\sum_{n\in\mathbb{Z}}
\!\int_0^\infty\!\frac{dT}{T^{3/2}}
\!\int d\bar C_0\,d\xi\, dC_0\,d\eta\;
\exp\!\left[
-\frac{1}{2}\!\left(m^2 T + \frac{n^2\beta^2}{T}\right)
-\bar C_0\xi - C_0\eta 
\right]
\\[4pt]
=\frac{\beta}{2(2\pi)^{1/2}}
\sum_{n\in\mathbb{Z}}
\!\int_0^\infty\!\frac{dT}{T^{3/2}}
\!\int d\bar C_0\,d\xi\;
\exp\!\left[
-\frac{1}{2}\!\left(m^2 T + \frac{n^2\beta^2}{T}\right)
-\bar C_0\xi
\right].
\end{gathered}
\end{equation}
In going to the last line we have integrated out 
$(C_0,\eta,\rho)$ using the appropriately normalized Gaussian measure
$d\rho$.  
This reproduces the semiclassical expression~\eqref{eq:IntegralForm},
now keeping explicit the contributions from the ghost zero modes.

\subsection{Localization of the Modulus}
 Here we apply supersymmetric localization to 
\begin{equation}\label{eq:1stloc}
W(\beta)=-\frac{\beta}{2(2\pi)^{1/2}}
\sum_{n\in\mathbb{Z}}
\!\int_0^\infty\!\!\frac{dT}{T^{3/2}}
\int d\bar C_0\,d\xi\; e^{
-\frac{1}{2}\!\left(m^2T+\frac{n^2\beta^2}{T}\right)
-\bar C_0\xi},
\end{equation}
with the focus on the nonzero winding sectors $n\neq0$ (the $n=0$ mode only
shifts the vacuum energy, as discussed in
Section~\ref{sec:HOsemi}).

First, we observe that the exponent
\[
S_n(T)=\tfrac{1}{2}\!\left(m^2 T+\frac{n^2\beta^2}{T}\right)
\]
is invariant under the modular--type transformation
\[
T \;\longrightarrow\; \frac{|n|\beta}{mT}.
\]
This motivates rewriting the $n\neq 0$ part of \eqref{eq:1stloc} in a manifestly
symmetric form:
\begin{equation}
\begin{gathered}
W(\be)|_{n\neq 0} =-\;\frac{\beta}{4(2\pi)^{1/2}}\\
\times \sum_{n\in\mathbb{Z}\setminus\{0\}}
\int_0^\infty\!\frac{dT}{T}\,
\left(
\frac{1}{\sqrt{T}}
+\frac{m\sqrt{T}}{|n|\beta}
\right)\int d\bar C_0\,d\xi\; e^{
-\frac{1}{2}\!\left(m^2T+\frac{n^2\beta^2}{T}\right)
-\bar C_0\xi}.
\end{gathered}
\label{eq:modular-symmetric-form}
\end{equation}

To introduce the second layer supersymmetry, we extend the triplet $(T,\bar{C}_{0},\xi)$ by auxiliary bosonic field $E$ with appropriately normalized measure $dE$, so \eqref{eq:modular-symmetric-form} takes
the form
\begin{equation}
\begin{gathered}
W(\be)|_{n\neq 0} =-\;\frac{\beta}{4(2\pi)^{1/2}}
\sum_{n\in\mathbb{Z}\setminus\{0\}}
\int_0^\infty d\mu_{n}(T)\int d\bar C_0\,d\xi dE\; e^{-S_{n}(T,E,\bar{C}_{0},\xi)},
\end{gathered}
\label{eq:E-mode}
\end{equation}
where
\begin{equation}
\begin{gathered}
S_{n}(T,E,\bar{C}_{0},\xi)=\frac{1}{2}\,E^2 \;-\; iE\,f_{n}(T) \;+\; \bar C_0\,\xi \;+\; |n| \beta m,\\
f_{n}(T)=m\sqrt{T}-\frac{|n|\beta}{\sqrt{T}},\\
d\mu_{n}(T)=\frac{f_{n}'(T)}{n\beta}\,dT
=\frac{1}{2T}\!\left(
\frac{1}{\sqrt{T}}
+\frac{m\sqrt{T}}{|n|\beta}
\right)dT.
\end{gathered}
\end{equation}
The zero-dimensional multiplet $(T,\xi,\bar C_0,E)$ in the $n$-th sector carries the
off-shell supersymmetry
\[
\delta_0 T=\frac{\xi}{f_{n}'(T)},\qquad
\delta_0\xi=0,\qquad
\delta_0\bar C_0=iE,\qquad
\delta_0E=0.
\]
Note that
the supersymmetry
$\delta_0$ is well-defined since
$f_{n}'(T)\neq0$ for all $T>0$, so 
the transformation $\delta_0 T$ is regular on the
integration domain.

This structure precisely parallels the off-shell localization analysis of the
zero-dimensional version of Witten's Morse theory \cite{Witten:1982im} (or Landau-Ginzburg model),
discussed in Appendix~\ref{sec:0d-LG-toy}. 
Indeed, we have
\begin{equation}
-\,iE\,f_{n}(T)+\bar C_0\,\xi
=~
\delta_0\!\big(-\,\bar C_0\,f_{n}(T)\big),
\end{equation}
so 
$$S_{n}=\frac{1}{2}\,E^2 + |n| \beta m + \delta_0\!\big(-\,\bar C_0\,f_{n}(T)\big)$$
and we consider the deformation $V_{n}=-\,\bar C_0\,f_{n}(T)$. Then for all $\lambda_{0}$ we have
\begin{equation}\label{W-V1}
\begin{gathered}
W(\beta)=-\;\frac{\beta}{4(2\pi)^{1/2}}
\sum_{n\in\mathbb{Z}\setminus\{0\}}
\int_0^\infty d\mu_{n}(T)\int d\bar C_0\,d\xi dE\; e^{-S_{n}-\lambda_{0}\delta_{0}V_{n}},
\end{gathered}
\end{equation}
provided that the full integration measure 
$d\mu_{n}(T)\,d\bar C_0\,d\xi\,dE$ is $\delta_{0}$-invariant.

Indeed,  consider $\delta_0$ as a change of variables
\[
T' \;=\; T + \varepsilon\,\frac{\xi}{f_{n}'(T)},\qquad
\xi' \;=\; \xi,\qquad
\bar C_0' \;=\; \bar C_0 + i\,\varepsilon\,E,\qquad
E' \;=\; E
\]
with a Grassmann parameter $\varepsilon$. 
Since Berezin measure is translation invariant, $d\bar{C}_{0}^{\prime}=d\bar{C}_{0}$, and it is sufficient to check that $d\mu(T')=d\mu(T)$. 
We have
\[f_{n}'(T')=f'_{n}(T)+\varepsilon\,\frac{\xi\,f_{n}''(T)}{f_{n}'(T)},\qquad
dT' =
\left(1-\varepsilon\,\xi\,\frac{f_{n}''(T)}{f_{n}'(T)^2}\right)dT,
\]
so
\[
\frac{f_{n}'(T')}{n\beta}\,dT'
=\frac{f_{n}'(T)}{n\beta}\,dT.
\]

Now integrating out $E$ in \eqref{W-V1} gives
\begin{equation*}
\begin{gathered}
W(\beta)=-\;\frac{\beta}{4(2\pi)^{1/2}}
\sum_{n\in\mathbb{Z}\setminus\{0\}}e^{-|n| \beta m}
\int_0^\infty d\mu_{n}(T)e^{-\frac{\lambda_0^2}{2}\,f_{n}(T)^2}\int d\bar C_0\,d\xi e^{-\lambda_0\,\bar C_0\,\xi},
\end{gathered}
\end{equation*}
where it is assumed that $\mathrm{Re}\, \lambda_0^2>0$. Thus as $\lambda_{0}\to\infty$, the integral localizes at the critical points --- the zero set locus $f_{n}(T_{n})=0$, where $T_{n}=|n|\beta/m$.

The prefactor\footnote{Of course, after a change of variable the integral over $T$ is Gaussian.} of the localization over modulus $T$ is the  the standard zero-dimensional LG prefactor, which cancels the Grassmann integral
$$\!\int d\bar C_0\,d\xi\,e^{-\lambda_0\bar C_0\xi}=\lambda_0$$
and gives a $\lambda_0$-independent result:
\[ \lim_{\lambda_{0}\to\infty}\int_0^\infty d\mu_{n}(T)e^{-\frac{\lambda_0^2}{2}\,f_{n}(T)^2}\int d\bar C_0\,d\xi e^{-\lambda_0\,\bar C_0\,\xi}
=
\frac{\sqrt{2\pi}}{n\be}
  e^{-|n|\be m}\!.
\]
Substituting to \eqref{eq:modular-symmetric-form} finally gives
\begin{equation}
W(\be)|_{n\neq 0}
   = -\frac{1}{2}\!
   \sum_{n\neq0}\!\frac{e^{-|n|\beta m}}{|n|}
   = \log\! \left(1-e^{-\be m}\right),
   \end{equation}
which exhibits the exactness of the semiclassical analysis presented in
Section~\ref{sec:HOsemi}.

\section{Scalar QED and Schwinger Pair Production} \label{sec: Schwinger}

Consider scalar quantum electrodynamics on a fixed classical
electromagnetic background in Minkowski space~$\mathbb{R}^{3,1}$.
The vacuum--to--vacuum transition amplitude in the presence of a
background gauge field~$A_\mu(x)$ is
\begin{equation}
\langle 0_{\mathrm{out}}|\,0_{\mathrm{in}}\rangle_{A}
~=~
Z[A]
~=~
\int\!\mathscr D\phi\,\mathscr D\phi^{\ast}\;
\exp\!\big(i\,S_{\mathrm{L}}[\phi,\phi^{\ast};A]\big),
\label{eq:Z-Lorentz}
\end{equation}
where the Lorentzian scalar QED action is
\begin{equation}
S_{\mathrm{L}}[\phi,\phi^{\ast};A]
~=~
\int d^4x\,
\Big[
- (D_\mu\phi)^{\ast} D^\mu\phi
- m^2\,\phi^{\ast}\phi
\Big],
\qquad
D_\mu=\partial_\mu+i q A_\mu,
\label{eq:Lorentz-action}
\end{equation}
and the metric is taken to be $\eta_{\mu\nu}=\mathrm{diag}(-,+,+,+)$.

The one--loop effective action in the given background is defined by
\begin{equation}
Z[A]
~=~
\langle 0_{\mathrm{out}}|\,0_{\mathrm{in}}\rangle_{A}
~=~
\exp\!\big(i\,\Gamma[A]\big),
\label{eq:Gamma-def}
\end{equation}
so that $\Gamma[A]$ encodes the response of the quantum vacuum.
Its imaginary part determines the probability for vacuum decay via
pair production,
\begin{equation}
P_{\text{pair}}
~=~
1 - |\langle 0_{\mathrm{out}}|\,0_{\mathrm{in}}\rangle_{A}|^2
~=~
1 - e^{-2\,\mathrm{Im}\,\Gamma[A]}.
\label{eq:pair-prob}
\end{equation}

\subsection{Lorentzian Worldline Representation and Wick rotation}
\label{subsec:worldline-wick}

The one--loop effective action $\Gamma[A]$ can be written as a functional
determinant of the Klein--Gordon operator in the background $A_\mu$.
From \eqref{eq:Lorentz-action} and \eqref{eq:Gamma-def} we obtain
\begin{equation}
\Gamma[A]
~=~
-i\,\mathrm{Tr}\,\log\big(-D^2 - m^2 + i\epsilon\big),
\qquad
D_\mu = \partial_\mu + i q A_\mu,
\label{eq:Gamma-trace-log}
\end{equation}
where the $i\epsilon$ prescription selects the vacuum boundary
conditions.  The standard Schwinger proper--time representation is
\begin{equation}
\log\!\big(-D^2 - m^2 + i\epsilon\big)
~=~
- \int_0^\infty \frac{dT}{T}\,
\exp\!\big(-iT\,(-D^2 - m^2 + i\epsilon)\big).
\label{eq:schwinger-L-standard}
\end{equation}
As discussed in Section~\ref{sec:HOsemi}, the relation
\eqref{eq:schwinger-L-standard} should be understood as
\(\simeq\), since the right–hand side is UV divergent and requires
renormalization.  For notational convenience, we will nevertheless treat
it as an equality in the formulas below, and return to the required
renormalization in the Section \ref{instantons}.

For later convenience we rescale the proper time by 
\(
T \rightarrow T/2,
\)
so that the worldline kinetic term appears with coefficient
\(1/(2T)\).  This gives the equivalent form
\begin{equation}
\log\!\big(-D^2 - m^2 + i\epsilon\big)
~=~
- \int_0^\infty \frac{dT}{T}\,
\exp\!\Big(-\tfrac{iT}{2}\,(-D^2 - m^2 + i\epsilon)\Big),
\label{eq:schwinger-L}
\end{equation}
and hence
\begin{equation}
\Gamma[A]
~=~
i \int_0^\infty \frac{dT}{T}\,
e^{-i(m^2 - i\epsilon)T/2}\,
\mathrm{Tr}\,e^{\,iT D^2/2}.
\label{eq:Gamma-proper-time}
\end{equation}

The trace can be represented as a sum over closed trajectories of a
relativistic particle.  Writing the kernel
$K(x_f,x_i;T)=\langle x_f|e^{iT D^2/2}|x_i\rangle$, one has
\begin{equation}
\mathrm{Tr}\,e^{\,iT D^2/2}
~=~
\int d^4x\,K(x,x;T)
~=~
\int d^4x_0 \int_{x(0)=x(T)=x_0}\!\!\mathscr D x(\tau)\;
\exp\!\big(i S_{\mathrm{L}}[x;T]\big),
\label{eq:trace-worldline}
\end{equation}
where $\tau\in[0,T]$ parametrizes the closed loop.
It is convenient to rescale $\tau\to T\tau$ so that $\tau\in[0,1]$ and
\begin{equation}
S_{\mathrm{L}}[x;T]
~=~
\int_0^1 d\tau\,
\Big[
\frac{1}{2T}\,\dot x^2(\tau)
+ q\,\dot x^\mu(\tau)\,A_\mu(x(\tau))
\Big],
\qquad
\dot x^2 = \eta_{\mu\nu}\dot x^\mu\dot x^\nu.
\label{eq:worldline-L}
\end{equation}
The one--loop effective action then becomes
\begin{equation}
\Gamma[A]
~=~
i \int_0^\infty \frac{dT}{T}\,
e^{-i(m^2 - i\epsilon)T/2}
\int_{\mathrm{PBC}}\mathscr D x\;
e^{\,i S_{\mathrm{L}}[x;T]},
\label{eq:Gamma-worldline-L}
\end{equation}
where PBC stands for the periodic boundary conditions $x^\mu(1)=x^\mu(0)$.

\subsubsection*{Constant electric field and contour rotation}

Let us specialize to a constant electric field $E$ pointing in the
$x^1$--direction.  A convenient gauge choice is
\begin{equation}
A_1(x) = E\,x^0,
\qquad
A_0=A_2=A_3=0,
\label{eq:gauge-constant-E}
\end{equation}
so that $F_{01}=E$.  Then the interaction term in
\eqref{eq:worldline-L} simplifies to
\begin{equation}
S_{\mathrm{int,L}}[x]
~=~
qE \int_0^1 d\tau\,x^0(\tau)\,\dot x^1(\tau).
\label{eq:worldline-int-L}
\end{equation}
The Lorentzian worldline action becomes
\begin{equation}
S_{\mathrm{L}}[x;T]
~=~
\int_0^1 d\tau\,
\Big[
\frac{1}{2T}\,\big(-\dot x_0^2 + \dot x_1^2 + \dot x_\perp^2\big)
+ qE\,x^0\,\dot x^1
\Big],
\label{eq:worldline-L-constant-E}
\end{equation}
with $\dot x_\perp^2=\dot x_2^2+\dot x_3^2$.

To obtain the Euclidean representation it is useful to view the Wick
rotation as a contour deformation in complexified field space.
We leave $x^i(\tau)$, $i=1,2,3$, real and rotate the proper time $T$
and the time coordinate $x^0$ by the same angle $\theta$ in the lower
half--plane (an analogous rotation was used in the localization
proof of the Selberg trace formula \cite{Choi:2023pjn}),
\begin{equation}
T = e^{i\theta} T_{\mathrm{E}},
\qquad
x^0(\tau) = e^{i\theta} x^0_{\mathrm{E}}(\tau),
\qquad
\theta : 0 \to -\frac{\pi}{2}.
\label{eq:contour-rotation}
\end{equation}
For $-\tfrac{\pi}{2}<\theta<0$ the quadratic kinetic term acquires a
negative real part, ensuring convergence of the Gaussian functional
integral in each worldline mode.  The factor
$e^{-i(m^2-i\epsilon)T/2}$ in \eqref{eq:Gamma-worldline-L}
analogously guarantees convergence of the proper--time integral as its
contour is rotated.  By Cauchy's theorem the deformation may be
performed continuously, and the value of the integral is unchanged as
$\theta$ is taken from $0$ to $-\tfrac{\pi}{2}$.

Under \eqref{eq:contour-rotation} every piece of the Lorentzian action
transforms analytically, and at $\theta=-\tfrac{\pi}{2}$ one finds:
\begin{itemize}
\item the kinetic term becomes the usual positive--definite Euclidean
quadratic form,
\begin{equation}
i \int_0^1 d\tau\,
\frac{1}{2T}\,\big(-\dot x_0^2 + \dot x_1^2 + \dot x_\perp^2\big)
\;\longrightarrow\;
- \int_0^1 d\tau\,
\frac{1}{2T_{\mathrm{E}}}\,
\big(\dot x_{0,\mathrm{E}}^2 + \dot x_1^2 + \dot x_\perp^2\big);
\label{eq:kinetic-wick}
\end{equation}

\item the mass term becomes exponentially damped,
\begin{equation}
e^{-i(m^2-i\epsilon)T/2}
\;\longrightarrow\;
e^{-m^2 T_{\mathrm{E}}/2};
\label{eq:mass-wick}
\end{equation}

\item the minimal coupling contributes
\begin{equation}
i\,qE\!\int_0^1\! d\tau\,x^0\dot x^1
\;\longrightarrow\;
qE\!\int_0^1\! d\tau\,x^0_{\mathrm{E}} \dot x^1,
\end{equation}
so its Euclidean contribution is
\begin{equation}
S_{\mathrm{int,E}}[x_{\mathrm{E}}]
~=~
-\,qE \int_0^1 d\tau\,x^0_{\mathrm{E}}(\tau)\,\dot x^1(\tau).
\label{eq:int-wick}
\end{equation}
\end{itemize}

Collecting all terms and using $\cD x=-i\cD x_{\text{E}}$ yields the Euclidean worldline representation of
the effective action,
\begin{equation}
\Gamma[A]
~=~
\int_0^\infty \frac{dT_{\mathrm{E}}}{T_{\mathrm{E}}}\,
e^{-m^2 T_{\mathrm{E}}/2}
\int_{\mathrm{PBC}}\mathscr D x_{\mathrm{E}}\;
e^{- S_{\mathrm{E}}[x_{\mathrm{E}};T_{\mathrm{E}}]},
\label{eq:Gamma-worldline-E}
\end{equation}
with the Euclidean action
\begin{equation}
S_{\mathrm{E}}[x_{\mathrm{E}};T_{\mathrm{E}}]
~=~
\int_0^1 d\tau\,
\Big[
\frac{1}{2T_{\mathrm{E}}}\,
\big(\dot x_{0,\mathrm{E}}^2 + \dot x_1^2 + \dot x_\perp^2\big)
- qE\,x^0_{\mathrm{E}}\,\dot x^1
\Big].
\label{eq:worldline-E-constant-E}
\end{equation}

The Euclidean worldline action can be written covariantly as
(following standard conventions in the literature,
e.g.\ \cite{Affleck:1981bma,Dunne:2005sx})
\begin{equation}
S_{\text{world}}^{\text{E}}
= \int_{0}^{1} d\tau \left[
\frac{1}{2T}\,\dot x^\mu \dot x_\mu
+ \frac{T}{2}\,m^{2}
+ i q\,A_\mu(x(\tau))\,\dot x^\mu
\right],
\label{eq:worldline-E-start}
\end{equation}
where \(x^\mu(\tau)\in \mathbb{R}^{4}\).
From this point onward we suppress the Euclidean subscript and simply
write \(x^\mu\) for the Euclidean worldline coordinates.

Here the background potential is implicitly understood to depend on the
Wick-rotated time coordinate,
\[
A_\mu(x^{0},x^{i}) \;\longrightarrow\; A_\mu(i x^{0}, x^{i}),
\]
and we do not perform the tensorial continuation \(A_{0} \to i A_{0}\).
For the constant electric field of
Section~\ref{subsec:worldline-wick}, the Lorentzian gauge choice
\(A_1(x)=E\,x^0\) becomes \(A_1(x)=iEx^0\), with
\(A_0=A_2=A_3=0\).  This is the gauge potential that enters the
Euclidean worldline action \eqref{eq:worldline-E-start}.

This leads to the final Euclidean worldline representation of the 
effective action:
\begin{equation}
\Gamma[A]
=\int_{0}^{\infty} \frac{dT}{T}\;
\int_{\text{PBC}}\! \cD x(\tau)\;
\exp\!\left[
- \int_{0}^{1} d\tau \left(
\frac{1}{2T}\,\dot x^\mu \dot x_\mu
+ \frac{T}{2}\,m^{2}
+ i q\,A_\mu(x(\tau))\,\dot x^\mu
\right)
\right].
\label{eq:Schwinger-worldline-start}
\end{equation}

Finally, we note that the Euclidean proper–time contour should be taken
to lie slightly above the positive real axis,
\[
T \in \mathbb{R}_{>0} + i0^{+},
\]
as emphasized already in \cite{Schwinger:1951nm}.  
This follows directly from the contour deformation 
\(T = e^{i\theta} T_{\mathrm E}\) with 
\(\theta \to -\pi/2\) in \eqref{eq:contour-rotation}, which carries the
Lorentzian \(i\epsilon\) prescription into the Euclidean representation.
The slight upward displacement of the \(T\)-contour is therefore not an
additional assumption but the natural continuation of the original
Lorentzian boundary conditions, ensuring that no potential singularities
of the analytically continued integrand lie on the integration path.

\subsection{From Euler--Heisenberg to Worldline Instantons}
\label{subsec:EH-Schwinger-AAM}

Historically, the proper–time representation of the effective action in
a constant background was first written by Euler and Heisenberg
\cite{Heisenberg:1936nmg}, and its scalar analogue by Weisskopf
\cite{Weisskopf:1939zz}.  In these works the proper time integrals were
treated entirely in Euclidean signature and no imaginary part was
extracted; the vacuum instability of a constant electric field was not
yet understood.  The appearance of a nonzero imaginary part, signalling
pair production, was first recognized by Schwinger \cite{Schwinger:1951nm}.
Starting from the Lorentzian proper–time formula
\begin{equation}
\mathscr{L}(E)=\int_{0}^{\infty}\frac{dT}{T}\,
e^{-i(m^{2}-i\epsilon)T}\,f(T;E),
\end{equation}
he rotated the contour into the Euclidean direction in a manner
consistent with the Feynman $i\epsilon$ prescription.  The poles of
$f(T;E)$ that lie on the Lorentzian real axis then contribute through
their residues, yielding the vacuum decay rate.  For a charged scalar,
\begin{equation}
\operatorname{Im}\,\mathscr{L}_{\rm scalar}(E)
=\frac{(qE)^{2}}{8\pi^{3}}
\sum_{n=1}^{\infty}\frac{(-1)^{n+1}}{n^{2}}
\exp\!\left[-\,\frac{n\pi m^{2}}{qE}\right],
\end{equation}
the celebrated Schwinger formula; see
\cite{Dunne:2004nc} for a comprehensive review.

\medskip

A remarkable observation was made by Affleck, Alvarez, and Manton (AAM)
\cite{Affleck:1981bma}.  Working directly with the Euclidean worldline
path integral
\begin{equation}
\Gamma[A]
~=~
\int_0^\infty\!\frac{dT_{\mathrm E}}{T_{\mathrm E}}\,
e^{-m^{2} T_{\mathrm E}/2}
\int_{\mathrm{PBC}}\cD x\;
e^{-S_{\mathrm E}[x;T_{\mathrm E}]},
\end{equation}
they found that extremizing the combined functional
\(
S_{\mathrm E}[x;T_{\mathrm E}] + m^{2} T_{\mathrm E}/2
\)
with respect to both $x(\tau)$ and $T_{\mathrm E}$ produces a discrete
set of isolated critical point trajectories, even in this
reparametrization--invariant setting.

For a constant electric field the critical points take the form of
circular loops of radius $R = m/(qE)$ with winding number $n\ge1$ —
the “worldline instantons’’ \cite{Dunne:2005sx,Dunne:2006st}.  Their
appearance is highly nontrivial: the worldline equations admit a
discrete set of isolated Euclidean trajectories that dominate the
path integral.  In the gauge where the electric field points in the
$(x^0,x^1)$ direction, the instanton solutions lie entirely in this
plane and may be written explicitly as
\begin{equation}
\boxed{
\begin{aligned}
x^0_{\mathrm{cl}}(\tau) &= x^0_{\rm cm}
   + R \sin(2\pi n\,\tau),\\[3pt]
x^1_{\mathrm{cl}}(\tau) &= x^1_{\rm cm}
   + R \cos(2\pi n\,\tau),\\[3pt]
x^2_{\mathrm{cl}}(\tau) &= x^2_{\rm cm},\qquad
x^3_{\mathrm{cl}}(\tau) = x^3_{\rm cm},
\end{aligned}
}
\qquad 0\le\tau\le 1.
\end{equation}
Here $x^\mu_{\rm cm}$ is the center-of-mass zero mode.  Each
instanton possesses a single negative fluctuation mode — the radial
deformation of the circular loop — which yields the imaginary part of
$\Gamma$ through the standard semiclassical tunneling prescription.
This picture provides a clear geometric understanding of the Schwinger effect.

\medskip

That the AAM evaluation captures the Schwinger exponent so precisely
within a semiclassical framework naturally raises the question of
whether a localization-like structure may be present in the worldline
formalism.  An attempt along these lines was made by Gordon and Semenoff
\cite{Gordon:2014aba,Gordon:2016ldj}, based on the observation that the
circular instantons possess a rigid $S^{1}$ symmetry corresponding to
shifts of the worldline parameter.  By introducing a BRST symmetry associated with this mode, they argued
that fluctuations beyond one loop cancel.  While their
construction captures certain geometric features of the instanton
solutions, the derivation does not appear to yield a fully consistent
localization principle, and some conceptual steps remain unclear.

In the following section we take a different route, extending the
supersymmetries developed in the harmonic oscillator example to
construct a genuine localization framework for the worldline path
integral.  This approach localizes the full functional integral — both its
real and imaginary parts — and provides a controlled derivation of the
contributions to $\operatorname{Im}\Gamma$.

\section{Worldline Localization for the Schwinger Pair Production}
\label{sec:Schwingerloc}

Our starting point is the Euclidean worldline path integral for the 
effective action \(\Gamma[A]\) in 
\eqref{eq:Schwinger-worldline-start}.  
In this section we analyze the localization structure in general 
\(d\)-dimensional spacetime, keeping the background electric field fixed 
and applying the hidden fermionic symmetries developed earlier to the 
scalar QED worldline theory.

%\subsection{BRST/BV Structure with Multiple Reparametrizations}
%Before analyzing the BRST/BV structure, it is convenient to introduce a
%bookkeeping device that makes the localization structure transparent.
%Although the physical worldline is one–dimensional, the Euclidean
%worldline action for scalar QED in a constant electric field \eqref{eq:worldline-E-start} factorizes
%into a longitudinal $(0,1)$ sector — where the field acts — and $(d\!-\!2)$
%transverse free sectors:
%\[
%S_{\mathrm{world}}^{E}
%=
%S_{\parallel}[x^0,x^1;T]
%\;+\;
%\sum_{i=2}^{d-1} S_{\perp}[x_i;T].
%\]
%
%For localization purposes it is technically useful to let each sector
%carry its own copy of the worldline reparametrization symmetry.  This is
%implemented by introducing a set of dummy parameters
%\[
%\tau_\parallel\in[0,1],
%\qquad
%\tau_i\in[0,1]\quad (i=2,\dots,d-1),
%\]
%together with einbeins \(e_\parallel(\tau_\parallel)\) and \(e_i(\tau_i)\).
%All reparametrizations share the same proper–time modulus, imposed as
%the constraint
%\[
%T=\int_0^1 e_\parallel(\tau_\parallel)\,d\tau_\parallel
%    =\int_0^1 e_i(\tau_i)\,d\tau_i .
%\]
%
%This rewriting does not introduce new physical degrees of freedom.
%Rather, it makes explicit that each factor in the action has its own
%one–dimensional diffeomorphism symmetry, allowing the BRST/BV machinery
%to be applied independently in the longitudinal and transverse sectors.
%After gauge fixing the system reduces back to the standard single–modulus
%worldline representation with measure \(dT/T\).
%
%With this structure in place, we now present the corresponding BV
%gauge fixing.
%
%

\subsection{BRST/BV Structure with Multiple Reparametrizations}

Before analyzing the BRST/BV structure, it is convenient to introduce a
bookkeeping device that makes the localization structure transparent.
Although the physical worldline is one–dimensional, the Euclidean
worldline action for scalar QED in a constant electric field
\eqref{eq:worldline-E-start} factorizes into a longitudinal $(0,1)$
sector — where the field acts — and $(d\!-\!2)$ transverse free sectors:
\[
S_{\mathrm{world}}^{E}
=
S_{\parallel}[x^0,x^1;T]
\;+\;
\sum_{i=2}^{d-1} S_{\perp}[x_i;T].
\]

For localization purposes it is technically useful to let each sector
carry its own copy of the worldline reparametrization symmetry.  
This is implemented by introducing dummy worldline parameters
\[
\tau_\parallel\in[0,1],
\qquad
\tau_i\in[0,1]\quad (i=2,\dots,d-1),
\]
together with einbeins \(e_\parallel(\tau_\parallel)\) and
\(e_i(\tau_i)\), subject to the constraint that all sectors share the
same proper–time modulus:
\[
T=\int_0^1 e_\parallel(\tau_\parallel)\,d\tau_\parallel
    =\int_0^1 e_i(\tau_i)\,d\tau_i .
\]

This rewriting introduces no new physical degrees of freedom; it merely
makes explicit that each sector carries a one–dimensional
diffeomorphism symmetry.  
This allows the BRST/BV machinery to be applied independently in the
longitudinal and transverse sectors.  
After gauge fixing the theory reduces to the usual single–modulus
representation with measure \(dT/T\).

\medskip

The BRST/BV quantization proceeds exactly as in
Section~\ref{sec:BRST}, but with a separate ghost system for each
sector.  
We introduce
\[
\text{longitudinal ghost } C(\tau_\parallel), \qquad
\text{transverse ghosts } C_i(\tau_i),\qquad i=2,\dots,d-1,
\]
with corresponding antighosts
\(
\bar C(\tau_\parallel),\ \bar C_i(\tau_i)
\)
and constant zero–mode multiplets
\[
C_0,\ C_{i,0},\qquad
\bar C_0,\ \bar C_{i,0},\qquad
\xi,\ \xi_i,\qquad
\eta,\ \eta_i.
\]

After fixing the derivative gauges
\(
\dot e_\parallel = 0,\ \dot e_i = 0
\)
and reducing all einbeins to a single proper–time modulus
\(
 T,
\)
the gauge–fixed BRST/BV action takes the following decoupled form
\begin{equation}\label{eq:BV-multi-tau}
\boxed{
\begin{gathered}
S_{\mathrm{BRST}}^{\mathrm{gf,red}}=
\\[4pt]
\int_0^1\!d\tau_\parallel\!\left[
 \frac{1}{2T}\left(\dot x_0^2(\tau_\parallel)
                 +\dot x_1^2(\tau_\parallel)\right)
 + \frac{T}{2}\,m^2
 + i q\,A_\mu(x)\,\dot x^\mu(\tau_\parallel)
 - \dot{\bar C}(\tau_\parallel)\,\dot C(\tau_\parallel)
\right] +\bar{C}_0\,\xi + C_0\,\eta \\[4pt]
+ \sum_{i=2}^{d-1}\int_0^1\!d\tau_i\left[
 \frac{1}{2T}\,\dot x_i^2(\tau_i)
 + \frac{T}{2}\,m^2
 - \dot{\bar C}_i(\tau_i)\,\dot C_i(\tau_i)
\right]
 + \sum_{i=2}^{d-1}\bar C_{i,0}\,\xi_i
+ \sum_{i=2}^{d-1}C_{i,0}\,\eta_i
\\[4pt]
=S_{\mathrm{BRST},\parallel}^{\mathrm{gf,red}} + S_{\mathrm{BRST},\perp}^{\mathrm{gf,red}}\,.
\end{gathered}
}
\end{equation}

Thus the full worldline theory decomposes into a longitudinal \((0,1)\)
sector coupled to the background electric field, together with \((d\!-\!2)\)
independent transverse sectors, each equipped with its own BRST/BV ghost
structure. Each transverse sector is described by the action  \eqref{eq:Sgf-reduced}, where $x_{i}(\tau_{i})$ takes values in $\RR$ instead of $\RR/\beta\ZZ$.

\subsection{Fermionic  Symmetries}

Because the worldline theory now has 
one longitudinal reparametrization in the $(0,1)$ plane
and $(d-2)$ independent transverse reparametrizations, 
the corresponding  supersymmetries split as
\[
\delta_{v} \quad\text{acting in the $(0,1)$ plane}, 
\qquad
\delta_i \quad\text{acting on } x_i,\qquad i=2,\dots,d-1.
\]

We introduce auxiliary fields
\[
G^m, \qquad m=0,1\qquad\text{and}\qquad 
G_i, \qquad i=2,\dots,d-1,
\]
and define the transformations as follows.

\paragraph{Longitudinal SUSY (two-dimensional).}
For a constant two–vector \(v^m\in\mathbb{R}^2\),
\begin{equation}
\boxed{
\begin{aligned}
\delta_v x^m
&= T^{3/2}\,v^m\,\dot{\bar C},\\[4pt]
\delta_v \dot C
&= T^{1/2}\,v_m\,\ddot x^m
 + iT\,v_m\,\dot G^m
 - i q\,T^{3/2}\,v_m F^{m}{}_{n}(x)\,\dot x^n,\\[4pt]
\delta_v \dot{\bar C} &= 0,\\[4pt]
\delta_v G^m &= iT\,v^m\,\ddot{\bar C}.
\end{aligned}
}
\label{eq:first-layer-longitudinal}
\end{equation}

\paragraph{Transverse SUSY for $x_i$, $i=2,\dots,d-1$.}
These directions are free and carry standard harmonic oscillator structure.
Their SUSY is
\begin{equation}
\boxed{
\begin{aligned}
\delta_i x_i &= T^{3/2}\,\dot{\bar C}_i,\\[4pt]
\delta_i \dot C_i &= T^{1/2}\,\ddot x_i + i T \dot G_i,\\[4pt]
\delta_i \dot{\bar C}_i &= 0,\\[4pt]
\delta_i G_i &= i T\ddot{\bar C}_i~.
\end{aligned}}
\label{eq:transverse-susy}
\end{equation}

\subsection{Supersymmetry Algebra}

The longitudinal  supersymmetry is parametrized by a constant
two–vector $v^m$, $m=0,1$, acting as in~\eqref{eq:first-layer-longitudinal}.
Given two such parameters $v_1^m$ and $v_2^m$, a direct computation shows that,
for a general background field $F_{\mu\nu}(x)$,
\begin{equation}
\boxed{
\begin{aligned}
\{\delta_{v_1},\delta_{v_2}\}\,\dot C
&= i q\,T^{3}\Big[
  (v_2\!\cdot\!\partial)F_{mn}\,v_1^m
 + (v_1\!\cdot\!\partial)F_{mn}\,v_2^m
 \Big]\dot x^n\,\dot{\bar C},
\end{aligned}
}
\label{eq:longitudinal-algebra-general}
\end{equation}
while
\begin{equation}
\{\delta_{v_1},\delta_{v_2}\}\,x^m
~=~\{\delta_{v_1},\delta_{v_2}\}\,\dot{\bar C}
~=~\{\delta_{v_1},\delta_{v_2}\}\,G^m
~=~0~.
\end{equation}
Here we use the shorthand
\[
(v F w)\;:=\;v^m F_{mn} w^n, 
\qquad
(v\!\cdot\!\partial)F_{mn}
\;:=\;v^p\,\partial_p F_{mn},
\qquad m,n,p=0,1.
\]

For a \emph{constant} field strength $F_{\mu\nu}=\text{const}$ (as in Schwinger's
setup), the derivative terms vanish and the longitudinal algebra reduces to
\begin{equation}
\boxed{
\{\delta_{v_1},\delta_{v_2}\}\,\dot C
=0.
}
\label{eq:longitudinal-algebra-constant}
\end{equation}
In particular, setting $v_1=v_2=v$ then implies
\[
\delta_v^{\,2} = 0
\quad\text{on all longitudinal fields},
\]
so each fixed $v$ generates an \emph{off–shell nilpotent} SUSY in
the $(0,1)$ plane for constant $F$.

\medskip

In the transverse sectors, each $\delta_i$ acts only on 
$(x_i,\dot C_i, \dot {\bar C}_i,G_i)$ as in~\eqref{eq:delta1_offshell}. 
The algebra there is the same as in the single–reparametrization spinless case:
\begin{equation}
\delta_i^{\,2}=0,
\qquad
\{\delta_i,\delta_j\}=0\quad (i\neq j),
\end{equation}
since different sectors have disjoint sets of fields and no
background couplings.

Moreover, the longitudinal and transverse supersymmetries commute:
\begin{equation}
\{\delta_v,\delta_i\}=0,
\qquad \forall\,i=2,\dots,d-1,
\end{equation}
because $\delta_v$ acts only on $(x^m, \dot C, \dot{\bar C},G^m)$ and $\delta_i$ only on
$(x_i, \dot C_i,\dot {\bar C}_i,G_i)$.

Thus, for constant field strength, the system possesses
\[
d-1 = 1 + (d-2)
\]
mutually commuting, off–shell nilpotent first-layer supersymmetries: one
longitudinal family $\delta_v$ in the $(0,1)$ plane and $(d-2)$ transverse
supersymmetries $\delta_i$.

\subsection{Localizing Deformation}

We now specialize to a constant Euclidean electric field in the
$(0,1)$–plane,
\[
F_{01}=-F_{10}=iE, \qquad F_{\mu\nu}=0 \ \text{otherwise.}
\]
Introduce the complex polarization vectors
\[
u^\mu=(1,i,0,0),\qquad 
\bar u^\mu=(1,-i,0,0),
\]
and the associated complex coordinates
\[
z := x^0+i\,x^1, \qquad 
\bar z := x^0-i\,x^1, \qquad
G_z := G^0+i\,G^1, \qquad
G_{\bar z} := G^0-i\,G^1.
\]
In this basis the longitudinal SUSY acts on $\dot C$ as
\begin{equation}
\begin{aligned}
\delta_{u}\dot C
&= T^{\tfrac12}\,\ddot z
 + iT\,\dot G_z
 - i q E\,T^{\tfrac32}\,\dot z
 \;=\; T^{\tfrac12} K(z) + iT\,\dot G_z, \\[4pt]
\delta_{\bar u}\dot C
&= T^{\tfrac12}\,\ddot{\bar z}
 + iT\,\dot G_{\bar z}
 + i q E\,T^{\tfrac32}\,\dot{\bar z}
 \;=\; T^{\tfrac12} \bar K(z) + iT\,\dot G_{\bar z},
\end{aligned}
\label{eq:deltaC-complexpol-E}
\end{equation}
where we have introduced the rescaled combinations
\begin{equation}
K(z) := \ddot z - i q E\,T\,\dot z,
\qquad
\bar K(z) := \ddot{\bar z} + i q E\,T\,\dot{\bar z}.
\label{eq:K-def}
\end{equation}
Thus $K(z)=\bar K(z)=0$ are precisely the (Euclidean) cyclotron equations of
motion in the $(0,1)$–plane.

For the transverse directions $x_i$ with $i=2,\dots,d-1$ we use the
independent supersymmetries $\delta_i$, each with its own
ghost–auxiliary system $(C_i,\bar C_i,G_i)$, as constructed in the
multi–reparametrization setup.

We now choose the functionals
\begin{equation}
V_u
:= \int_0^1\!d\tau\;\bar K(z)\,C(\tau),
\qquad
V_i
:= \int_0^1\!d\tau_i\;\ddot x_i(\tau_i)\,C_i(\tau_i),
\quad i=2,\dots,d-1,
\label{eq:V-long-trans}
\end{equation}
and define the combined $\delta$–exact deformation
\begin{equation}
S_{\mathrm{def}}^{(1)}(\lambda)
:= \lambda\Big(
\delta_{u} V_u + \sum_{i=2}^{d-1}\delta_{i} V_i\Big).
\label{eq:Sdef-firstlayer-full}
\end{equation}

In the constant--field background he three supersymmetries are mutually commuting and nilpotent: 

\[
\delta_{u}^{\,2}=0,\qquad 
\delta_i^{\,2}=0,\qquad
\{\delta_{u},\delta_i\}=0,\qquad
\{\delta_i,\delta_j\}=0,\quad i\neq j.
\]
Indeed, $\delta_u$ acts only on the longitudinal fields
$(z,\bar z,G_z,G_{\bar z},C,\bar C)$, while each $\delta_i$ acts only on 
the transverse sector $(x_i,G_i,C_i,\bar C_i)$.  
Since these two sets of fields are disjoint, the operators 
$\delta_u$ and $\delta_i$ trivially anticommute:
\[
\delta_u\delta_i=\delta_i\delta_u=0,
\qquad
\delta_i\delta_j=\delta_j\delta_i=0,\quad i\neq j.
\]
Consequently, the deformation
$S_{\mathrm{def}}^{(1)}(\lambda)$ is invariant under all of them:
\[
\delta_{u}\,S_{\mathrm{def}}^{(1)}(\lambda)
~=~0, \qquad
\delta_{i}\,S_{\mathrm{def}}^{(1)}(\lambda)
~=~0.
\]

After integrating out out the auxiliary fields
$G_z,G_{\bar z},G_i$ as in the spinless case, the full deformation splits into longitudinal and transverse parts,
\[
S_{\mathrm{def}}^{(1)}(\lambda)
= S_{\mathrm{def},\parallel}^{(1)}(\lambda)
+ S_{\mathrm{def},\perp}^{(1)}(\lambda),
\]
where 
\begin{equation}
\begin{aligned}
S_{\mathrm{def},\parallel}^{(1)}(\lambda)
~=~&
\;\frac{\lambda}{T}\ \int_0^1 d\tau\,\bigl|K(z)\bigr|^2 + \lambda \int_0^1 d\tau\,
\bigl(\dddot{\bar C}(\tau) + i q E T\,\ddot{\bar C}(\tau)\bigr)\dot{C}(\tau),
%\\[4pt]
%&\;+\;
%\lambda \sum_{i=2}^{d-1} \int_0^1 d\tau_i\,
%\frac{1}{T}\,\bigl(\ddot x_i(\tau_i)\bigr)^2
%\;+\;
%\frac{\lambda^{2}}{2} \sum_{i=2}^{d-1} \int_0^1 d\tau_i\,
%\frac{1}{T}\,\bigl(\dddot x_i(\tau_i)\bigr)^2,
\end{aligned}
\label{eq:Sdef-firstlayer-||}
\end{equation}
%with bosonic part
\begin{equation}
\begin{aligned}
S_{\mathrm{def},\perp}^{(1)}(\lambda)
~=~&
%\;\lambda \int_0^1 d\tau\,\frac{1}{T}\,\bigl|K(z)\bigr|^2
%\\[4pt]
%&\;+\;
\frac{\lambda}{T}\ \sum_{i=2}^{d-1} \int_0^1 d\tau_i\,
\bigl(\ddot x_i(\tau_i)\bigr)^2
\;+\;
\frac{\lambda^{2}}{2T} \sum_{i=2}^{d-1} \int_0^1 d\tau_i\,
\bigl(\dddot x_i(\tau_i)\bigr)^2
\\[4pt]
&\;+\;
\lambda \sum_{i=2}^{d-1} \int_0^1 d\tau_i\,
\dddot{\bar C}_i(\tau_i)\,\dot C_i(\tau_i)\,,
\end{aligned}
\label{eq:Sdef-firstlayer-transverse}
\end{equation}
and
\[
K(z) := \ddot z - i q E T\,\dot z, \qquad z = x^0 + i x^1 .
\]

%and
%\begin{equation}
%\begin{aligned}
%S_{\mathrm{def}}^{(1)}(\lambda)\Big|_{\text{bos}}
%~=~&
%\;\lambda \int_0^1 d\tau\,\frac{1}{T}\,\bigl|K(z)\bigr|^2
%\\[4pt]
%&\;+\;
%\lambda \sum_{i=2}^{d-1} \int_0^1 d\tau_i\,
%\frac{1}{T}\,\bigl(\ddot x_i(\tau_i)\bigr)^2
%\;+\;
%\frac{\lambda^{2}}{2} \sum_{i=2}^{d-1} \int_0^1 d\tau_i\,
%\frac{1}{T}\,\bigl(\dddot x_i(\tau_i)\bigr)^2,
%\end{aligned}
%\label{eq:Sdef-firstlayer-transvers}
%\end{equation}
%The longitudinal fermionic piece is
%\begin{equation}
%S_{\mathrm{def},\parallel}^{(1)}(\lambda)\Big|_{\text{ferm}}
%~=~
%\lambda \int_0^1 d\tau\,
%\bigl(\dddot{\bar C}(\tau) + i q E T\,\ddot{\bar C}(\tau)\bigr)\,
%\dot C(\tau).
%\end{equation}
%
%For the transverse directions, the fermionic part has the same
%structure as in the harmonic oscillator case,
%\begin{equation}
%S_{\mathrm{def},\perp}^{(1)}(\lambda)\Big|_{\text{ferm}}
%~=~
%\lambda \sum_{i=2}^{d-1} \int_0^1 d\tau_i\,
%\dddot{\bar C}_i(\tau_i)\,\dot C_i(\tau_i)\,.
%\end{equation}

Thus we see that longitudinal and transverse directions localize independently and transverse deformation \eqref{eq:Sdef-firstlayer-transverse} is the sum of $d-2$ copies of the deformation \eqref{eq:Sdef-final-nosign}. 

To summarized, we have a representation 
\begin{equation}\label{Gamma-susy}
\begin{gathered}
\Gamma(A)=\lim_{\lambda\to\infty}\int_{0}^{\infty}\frac{dT}{T}\int\cD\bm{x}\cD\bm{\bar{C}}\cD\bm{C}\cD\bm{\xi}\cD\bm{\eta}\;e^{-S_{\mathrm{BRST}}^{\mathrm{gf,red}}-S_{\mathrm{def}}^{(1)}(\lambda)}\\[4pt]
=\lim_{\lambda\to\infty}\int_{0}^{\infty}\frac{dT}{T}\int\cD z\cD\bar{z}\cD\bar{C} \cD Cd\xi d\eta\;e^{-S_{\mathrm{BRST},\parallel}^{\mathrm{gf,red}}-S_{\mathrm{def},\parallel}^{(1)} (\lambda)}\\[4pt]
\times \lim_{\lambda\to\infty}\int_{0}^{\infty}\frac{dT}{T}\int\cD\bm{x}_{\perp}\cD\bm{\bar{C}}_{\perp} \cD\bm{C}_{\perp} \cD\bm{\xi}_{\perp}\cD\bm{\eta}_{\perp}\;e^{-S_{\mathrm{BRST},\perp}^{\mathrm{gf,red}}-S_{\mathrm{def},\perp}^{(1)}(\lambda)}\,,
\end{gathered}
\end{equation}
where we introduced a short-hand notations $\cD\bm{x}=\cD x_0\cdots\cD x_{d-1}$, $\cD\bm{x}_{\perp}=\cD x_2\cdots\cD x_{d-1}$, etc., to represent integration measure over all bosonic and fermionic longitudinal and transverse variables.
\subsection{Localization in the Transverse and Longitudinal Sectors}\label{sec:loc-long-transerse}
%The bosonic deformation \eqref{eq:Sdef-firstlayer-bos} forces
%\[
%K(z)=0, \qquad \ddot x_i(\tau_i)=0 \quad (i=2,\dots,d-1),
%\]
%so the longitudinal and transverse directions localize independently.
%We first consider the $(d-2)$ transverse coordinates $x^i$.
The second factor in \eqref{Gamma-susy} is the localization of the transverse path integral. The only difference with the analysis in Section is that each variable $x_{i,0}$ takes values in $\RR$ instead of the thermal circle. Therefore, it is simply
%
%--- localizes to the constant locus $x_i(\tau_i)=x_{i,0}$ and exactly as localization in the sector $n=-0$ in Section, 
%
%It follows from \eqref{eq:Sdef-firstlayer-||} that transverse modes localize on the solutions of the equations of motion
%\[
%\ddot x_i(\tau_i)=0,
%\]
%so the localization locus is just
%\[
%x_i(\tau_i)=x_{i,0}\qquad\text{(constant)}.
%\]
%This is precisely analogous to the $n=0$ sector in the harmonic oscillator
%analysis, except that here each $x_{i,0}$ takes values in $\mathbb{R}$.
%
%The Gaussian integration over the transverse oscillators, together with the
%fermionic partners and their zero modes, produces the standard free–particle
%prefactor.  Since each transverse direction contributes
%\(
%(2\pi T)^{-1/2}
%\)
%and the zero modes give only a trivial normalization, the full transverse
%contribution is
\begin{equation}
%\int \mathscr{D}(\text{transverse sector})
%\;\longrightarrow\;
 \lim_{\lambda\to\infty}\int\cD\bm{x}_{\perp}\cD\bm{\bar{C}}_{\perp}\cD\bm{C}_{\perp}\cD\bm{\xi}_{\perp}\cD\bm{\eta}_{\perp}\;e^{-S_{\mathrm{BRST},\perp}^{\mathrm{gf,red}}-S_{\mathrm{def},\perp}^{(1)}(\lambda)} = 
\frac{\mathrm{Vol}(\mathbb{R}^{\,d-2})}
     {(2\pi T)^{(d-2)/2}}\, .
\label{eq:transverse-factor}
\end{equation}

\subsubsection{Localization of the Longitudinal Sector}

The longitudinal localization equation
\[
K(z)\;=\;\ddot z - i q E T\,\dot z \;=\; 0
\]
is a first–order linear ODE for $\dot z(\tau)$.  A convenient way to
parametrize its solution is
\begin{equation}
z(\tau)
~=~ z_0 + R\,e^{\,i\omega T\,\tau},
\qquad
\omega \equiv qE,
\label{eq:z-solution-general}
\end{equation}
with $z_0\in\mathbb C$ the center of the orbit and $R\in\mathbb C$ an
arbitrary complex parameter.  Thus the localized longitudinal trajectories
are Euclidean cyclotron orbits of radius \(|R|\), and the center-of-mass
position $z_0$ together with the complex modulus $R$ parametrizes the orbit
radius and its phase.

\medskip

Periodic boundary conditions impose
\[
z(1)=z(0)
\quad\Longleftrightarrow\quad
e^{\,i\omega T}=1.
\]
For generic $T$ this forces $R=0$, so the only localized configuration is
the trivial constant map $z(\tau)\equiv z_0$.

\medskip

However, whenever $T$ hits one of the discrete resonant values
\begin{equation}
T = T_n := \frac{2\pi n}{qE},
\qquad n\in\mathbb{Z}_{>0},
\label{eq:Tn-resonance}
\end{equation}
the phase factor satisfies $e^{\,i\omega T_n}=1$, and a whole circle of
nontrivial classical solutions survives:
\begin{equation}
z_n(\tau)
=
z_0 + R\,e^ {2\pi in\tau},
\qquad R\in\mathbb{C}.
\label{eq:zn-classical}
\end{equation}
We claim that these are the worldline instantons responsible for the Schwinger
pair–production amplitude.  At each resonant value $T=T_n$, the critical point is
highly degenerate: the solution space is parametrized by the complex
center-of-mass coordinate $z_0\in\mathbb C$ and the complex modulus
$R\in\mathbb C$, so that the family of instantons has the structure
$\mathbb C\times\mathbb C$.

%%%%%

\subsection{Trivial Longitudinal Saddles and the Euler--Heisenberg Effective Action}
%\subsubsection{Bosonic sector}
For each resonant point \(T_n = 2\pi n/(qE)\), it is convenient to introduce
 a small open neighborhood
\begin{equation}
U_n \equiv (\,T_n - \Delta T_n,\; T_n + \Delta T_n\,),
\qquad 
\frac{\Delta T_n}{T_n} \ll \frac{1}{\lambda},
\label{eq:Un-def}
\end{equation}
chosen small enough that different \(U_n\) do not overlap.  
We also define
\begin{equation}
\mathcal{U} \;\equiv\; \bigcup_{n\ge1} U_n,
\qquad
\mathcal{U}^{\,c} \;\equiv\; (0,\infty)\setminus\mathcal{U},
\label{eq:U-Uc-def}
\end{equation}
so that the proper--time axis decomposes as
\begin{equation}
(0,\infty)
\;=\;
\mathcal{U}\;\cup\;\mathcal{U}^{\,c}.
\label{eq:T-decomposition}
\end{equation}

For fixed $T\in \mathcal{U}^{\,c}$ the localization equation admits
%For generic values of the proper time \(T\) which are \emph{not} of the
%resonant form \(T=T_n=2\pi n/(qE)\), the localization equation
%\[
%K(z)=\ddot z - i q E T\,\dot z = 0
%\]
%with periodic boundary conditions admits only the trivial solution
%\[
%z(\tau) = z_0 = \text{const}.
%\]
%
%
%In the complementary region \(\mathcal{U}^{\,c}\), the proper time is
%non-resonant, and the localization equation admits 
only trivial
constant solution \(z(\tau)=z_0\). Thus we consider 
\[
z(\tau) = z_0 + \delta z(\tau),
\]
with periodic \(\delta z(\tau)\) satisfying  
\[\int_0^1 d\tau\,\delta z(\tau)=0,\]
so  \(z_0\) is the only zero mode. 
%We therefore expand the deformation around this constant map.

Denote by $L_{T}=-\partial_\tau^2 + i q E T\,\partial_\tau$ the self-adjoint operator on $[0,1]$ with periodic boundary conditions. Its eigenvalues are $\lambda_{k}(T)=2\pi k(2\pi k-qET)$, $k\in\ZZ$.  Let 
$$\text{det}' L_{T}=\prod^{\infty}_{\substack{k=-\infty\\ k\neq 0}}\lambda_{k}(T)$$ 
be its zeta regularized determinant. It is given by the textbook formula
\begin{equation}\label{det-L}
\text{det}^{\prime}L_{T}=\frac{2}{qET}\sin\frac{qET}{2}.
\end{equation}
Correspondingly, the bosonic path integral in  \eqref{eq:Sdef-firstlayer-||} is
\begin{equation}\label{bose-gauss-generic}
\int \cD\delta z\cD\overline{\delta z} e^{-\frac{\lambda}{T}\int_{0}^{1}d\tau |L_{T}(\delta z)|^{2}}=\left(\frac{2\pi T}{\lambda}\right)^{-1}\frac{1}{(\det'L_{T})^{2}}.
\end{equation}
As in  \cite{Choi:2021yuz}, the prefactor in \eqref{bose-gauss-generic} comes from the fact that corresponding quantum dimension of
the space of fields $\delta z(\tau)$ is $2\zeta(0)=-1$.

Of course, \eqref{bose-gauss-generic} also can be obtained directly by performing the Gaussian integration using Fourier modes. Namely, put
\[\delta z=\sum^{\infty}_{\substack{k=-\infty\\k\neq 0}}c_{k}e^{2\pi ik\tau}.\]
Then
\begin{equation}
\begin{gathered}
\int \cD\delta z\,\cD\overline{\delta z}\;
e^{-\frac{\lambda}{T}\int_{0}^{1}d\tau\, |L_{T}(\delta z)|^{2}}
=
\prod^{\infty}_{\substack{k=-\infty\\k\neq 0}}
\int_{\CC} i\,dc_{k}\wedge d\bar{c}_{k}\;
e^{-\frac{\lambda}{T}\lambda_{k}(T)^{2}|c_{k}|^{2}}
\\[6pt]
=
\prod^{\infty}_{\substack{k=-\infty\\k\neq 0}}
\left(\frac{2\pi T}{\lambda}\right)\frac{1}{\lambda_{k}(T)^{2}}
=
\left(\frac{2\pi T}{\lambda}\right)^{-1}
\frac{1}{(\det' L_{T})^{2}}
\end{gathered}
\end{equation}
since $\prod_{n=1}^{\infty}a^{2}=e^{2\zeta(0)\log a}=1/a$ by the zeta function regularization.

Similarly, for the fermionic path integral in  \eqref{eq:Sdef-firstlayer-||} we get
\begin{equation}\label{fermion-general}
\int \cD \bar{C}_{\mathrm{osc}}\cD C_{\mathrm{osc}}\, e^{-\lambda \int_0^1 d\tau\,
\bigl(\dddot{\bar C}(\tau) + i q E T\,\ddot{\bar C}(\tau)\bigr)\dot{C}(\tau)}=\lambda ^{-1}\text{det}'L_{T}
\end{equation}
since $\det'L_{0}=1$ and we used $\text{det}'L_{-T}=\text{det}'L_{T}$. Of course, in terms of Fourier modes
\[
C(\tau)=\sum_{\substack{k=-\infty\\k\neq 0}}^{\infty}C_{k}e^{2\pi i k\tau}
\qquad\text{and}\qquad
\bar{C}(\tau)=\sum_{\substack{k=-\infty\\k\neq 0}}^{\infty}\bar{C}_{k}e^{2\pi i k\tau}
\]
the fermionic path integral \eqref{fermion-general} is equal to
\beq
\prod^{\infty}_{\substack{k=-\infty\\k\neq 0}}
\int d\bar{C}_{-k}\,dC_{k}\;
e^{\lambda (2\pi k)^{2}\lambda_{-k}(-T)\,\bar{C}_{-k}C_{k}}
&=
\prod^{\infty}_{\substack{k=-\infty\\k\neq 0}}
\lambda (2\pi k)^{2}(-\lambda_{k}(T))
\\[4pt]
&=\lambda^{-1}\,\text{det}'L_{T}
\eeq
since $\lambda_{-k}(-T)=\lambda_{k}(T)$ and $\prod_{k=1}^{\infty}(2\pi k)^{2}=1$ by zeta–function regularization.

To summarize, for fixed $T\in \mathcal{U}^{\,c}$ the deformation $S_{\mathrm{def},\parallel}^{(1)} (\lambda)$ localizes the field $z$ to
classical solutions $z(\tau)=z_{0}$ and completely integrates
out the oscillatory ghost modes $C_{\mathrm{osc}}$ and
$\bar C_{\mathrm{osc}}$.
Using formulas \eqref{bose-gauss-generic}-\eqref{fermion-general} we finally obtain
\begin{equation}\label{Gamma-long-EH}
\begin{gathered}
\lim_{\lambda\to\infty}\int\cD z\cD\bar{z}\cD\bar{C} \cD Cd\xi d\eta\;e^{-S_{\mathrm{BRST},\parallel}^{\mathrm{gf,red}}-S_{\mathrm{def},\parallel}^{(1)} (\lambda)} \\
=\text{vol}(\RR^{2})\times e^{-\frac{1}{2}m^{2}T}(2\pi T)^{-1}
\frac{\det'L_{T}\det'L_{0}}{(\det'L_{T})^{2}}\\
=\mathrm{Vol}(\mathbb{R}^2)\,
(2\pi T)^{-1}\, e^{-\frac{1}{2}m^{2}T}
\frac{\tfrac12\,qET}{\sin\!\left(\tfrac12\,qET\right)}.
\end{gathered}
\end{equation}

Combining the transverse factor \eqref{eq:transverse-factor} with the
longitudinal one--loop contribution %\footnote{Of course, one can also obtain this result by performing the Gaussian integration using Fourier modes.
\eqref{Gamma-long-EH}, the
contribution from the trivial longitudinal critical point is therefore
\begin{equation}
\Gamma_{\text{triv}}[E]
=
\mathrm{Vol}(\mathbb{R}^{\,d})
\int_{\mathcal{U}^c} \frac{dT}{T}\,
(2\pi T)^{-d/2}\,
e^{-\tfrac12 m^2 T}\,
\frac{\tfrac12\,qET}{\sin(\tfrac12\,qET)},
\label{eq:Gamma-triv-localized}
\end{equation}
which gives the scalar Euler-Heisenberg integrand
in $d$ dimensions for a constant electric field.

%%%%%%%
\subsection{Instantons and the Schwinger Pair Production}\label{instantons}

As we have seen in Section \ref{sec:loc-long-transerse},
for the resonant proper times 
\begin{equation}
T = T_n = \frac{2\pi n}{qE},
\qquad n\in\mathbb{Z}_{>0},
\end{equation}
the longitudinal localization equation $K(z)=0$ admits a continuous family of classical solutions
\begin{equation}
z_n(\tau)
= z_0 + R\,e^{2\pi in\tau},
\qquad (z_0,R)\in\mathbb{C}\times\mathbb{C}
\label{eq:zn-moduli}
\end{equation}
--- Euclidean cyclotron orbits of winding number $n$, with $z_0$
the complex center of mass coordinate and $R$ the complex radius/phase
modulus.

A crucial fact --- which follows by inserting \eqref{eq:zn-moduli} into the
original worldline action \eqref{eq:worldline-E-constant-E} and using $qE\,T_n = 2\pi n$ --- is that  
\emph{the classical action is completely independent of the moduli:}
\begin{equation}
S_{\text{E}}[z_n,T_n] +\frac{m^{2}}{2}T_{n}
=
\frac{m^2}{2}\,T_n
=
\frac{\pi m^2}{qE}\,n.
%\qquad
%\text{independent of } (z_0,R).
\label{eq:Scl-independence}
\end{equation}
Thus the entire instanton manifold $\mathbb{C}\times\mathbb{C}$ lies at
the same classical action value: the instanton contribution is not a sum
over isolated critical points, but a true moduli integral.

Because of this degeneracy, the path integral cannot be treated by a
naive Gaussian expansion.  Instead, near each resonant point 
\(T_n = 2\pi n/(qE)\) we focus on the small neighborhood 
\(U_n = (T_n - \Delta T_n,\, T_n + \Delta T_n)\) introduced in
\eqref{eq:Un-def}--\eqref{eq:T-decomposition}.  
Within such a window we write
\[
T = T_n + \delta T, \qquad \delta T \in (-\Delta T_n,\, \Delta T_n),
\]
with $\Delta T_n \ll T_n$, and expand both the bosonic and fermionic 
deformation terms around the instanton manifold \eqref{eq:zn-moduli}.  
This local analysis reveals a distributional contribution supported at 
\(T=T_n\), which gives rise to the imaginary part of the effective action—the 
Schwinger pair-production rate of scalar QED.

Namely, we consider first the bosonic part in the deformation \eqref{eq:Sdef-firstlayer-||} and use a decomposition 
\(
z(\tau) = z_n(\tau) + \delta z(\tau),\;T = T_n + \delta T
\)
with periodic \(\delta z(\tau)\) satisfying  
\beq
\int_0^1 d\tau\,  z_n^* \delta z(\tau)=0.\label{ort-delta} \eeq
Using the expansion 
\[
K(z)=\ddot z - i qE\,T\,\dot z=-L_n\,\delta z - i qE\,\delta T\,\dot z_n + O(\delta^2),
\]
where
\(L_n=L_{T_{n}}=-\partial_\tau^2 + 2\pi i n\,\partial_\tau,
\)
for quadratic fluctuations we readily obtain
\[
\frac{\lambda}{T}\int_{0}^{1}d\tau |K(z)|^{2}=\frac{\lambda}{T_{n}}\int_{0}^{1}d\tau |L_{n}(\delta z)|^{2}+\lambda\alpha_{n}|R|^{2}(\delta T)^{2},\qquad \alpha_{n}=\frac{(2\pi n)^4}{T_n^3}.
\]
The subspace \eqref{ort-delta} of fluctuations $\delta z$ is orthogonal to zero modes $z_{n}$ of the operator $L_{n}$, so the regularized determinant of $L_{n}$ is a readily computed from \eqref{det-L} by removing the eigenvalue $2\pi n(2\pi n-qET)$ and passing to the limit $T\to T_{n}$:
%\begin{equation}\label{det-L-n}
%\det L_{n}=\lim_{T\to T_{n}}\frac{\det' L_{T}}{2\pi n(2\pi n-qET)}=\frac{(-1)^{n-1}}{(2\pi n)^{2}}.
%\end{equation}

Thus the bosonic path integral over $\cD\delta z$ is
\beq\label{bose-T-n}
\lim_{\lam\rightarrow \infty}\int\cD z\cD\bar{z} e^{-S_E(z,T)-\frac{\lambda}{T}\int_{0}^{1}d\tau |K(z)|^{2}}&=\lim_{T\to T_{n}} {  \left(\frac{2\pi T}{\lambda}\right)^{-1} \lam_n(T)^2 \ov \left(\frac{2\pi T}{\lambda}\right)({\det}'L_{T})^{2}  }  e^{-S_E[z_n,T_n]} e^{-\lambda\alpha_{n}(\delta T)^{2}}
\\&={(2\pi)^2 n^4 \ov T_n^2} \lam^2   e^{-{\pi m^2 \ov qE}n}   e^{-\lambda\alpha_{n}(\delta T)^{2}},
\eeq
so the modulis fluctation $\delta T$ gets `lifted'.

For the fermionic part, we again use Fourier modes
\[\bar{C}(\tau)=\sum_{k\in\ZZ}\bar{C}_{k}e^{2\pi ik\tau},\qquad C(\tau)=\sum_{k\in\ZZ} C_{k}e^{2\pi ik\tau},\]
so the deformation \eqref{eq:Sdef-firstlayer-||} takes the form
\[-\lambda\int_{0}^{1}d\tau L_{n}(\dot{\bar{C}})\dot{C} + \lambda qE(2\pi n)^{3}\delta T\bar{C}_{-n}C_{n},\]
where the first term does not contain instanton zero modes $\bar{C}_{n}, C_{n}$.
Correspondingly, we obtain 
\beq\label{fermi-loc}
\int \cD \bar{C}_{\mathrm{osc}} \cD C_{\mathrm{osc}}e^{\lambda\int_{0}^{1}d\tau L_{-n}(\dot{\bar{C}})\dot{C} - \lambda qE(2\pi n)^{3}\delta T\bar{C}_{-n}C_{n}}
&= \lim_{T\to T_{n}}  \frac{\lambda^{-1}\,\text{det}'L_{T} }{ -\lam (2\pi n)^{2} \lam_n(T)}\times  (\lambda qE(2\pi n)^{3}\delta T) \\
&={(-1)^{n}\ov (2\pi n )^4\lam^2} \times  (\lambda qE(2\pi n)^{3}\delta T)\\
&=(-1)^{n} {\del T \ov T_n  }
.
\eeq

The integration over zero modes $C_{0}, \bar{C}_{0}$ are taken care by the supersymmetry, and we are left with the Gaussian integral over the instanton moduli space
\begin{equation}\label{instanton-integral}
\int_{\CC\times\CC} d^{2}z_{0}d^{2}R e^{-\lambda\alpha_{n}(\delta T)^{2}|R|^{2}}=\text{vol}(\RR^{2})\times \frac{2\pi T_n^3 }{\lambda (2\pi n )^4(\delta T)^{2}}.
\end{equation}
%i\lambda(2\pi n)^{5}\deltaT \bar{c}_{-n}c_{n},\]
Putting together formulas \eqref{eq:transverse-factor}, \eqref{eq:Scl-independence}, \eqref{bose-T-n}, \eqref{fermi-loc} and \eqref{instanton-integral} together with the measure $dT/T$ in \eqref{Gamma-susy}, we obtain that in the localization limit $\lambda \to \infty$ the remaining
integration over the proper–time modulus in the neighborhood
\(
U_n = (T_n - \Delta T_n,\, T_n + \Delta T_n)
\)
with $\Delta T_n/T_n \ll 1/\lambda$ takes the form
\begin{align}
(-1)^{\,n}\,
\mathrm{Vol}(\mathbb R^d)\,
\frac{(qE)^{\,\frac d2}}{(2\pi)^{\,d}}\,
\frac{1}{n^{\,\frac d2}}\,
\exp\!\Bigl[-\,\frac{\pi m^{2}}{qE}\,n\Bigr]\,
\int_{T_n-\Delta T_n}^{T_n+\Delta T_n}{d(\delta T)\over \delta T}
\;,\qquad 
\Delta T_n\ll T_n.
\label{eq:deltaT-pole-integral-final}
\end{align}

With this structure in hand, we now
specify the contour for the proper ime integral.  As discussed in
Section~\ref{sec: Schwinger}, the physically correct Euclidean contour is
obtained by a small tilt
\begin{equation}
T \;\to\; T +i\varepsilon,
\qquad \varepsilon>0,
\end{equation}
so that near each resonant point $T=T_n$ the local variable
$\delta T = T-T_n$ is effectively replaced by $\delta T + i0^+$.
Using the distributional identity
\begin{equation}
\frac{1}{x+i0^+}
=\mathrm{v.p.}\,\frac{1}{x}
 - i\pi\,\delta(x),
\label{eq:PV-delta-identity}
\end{equation}
where $\mathrm{v.p.}$ denotes the Cauchy principal value, the integral
\[
\int_{-\Delta T_n}^{\Delta T_n}
\frac{d(\delta T)}{\delta T + i0^+}
\]
splits into a principal value part and a delta function contribution.
The principal value part combines with the trivial critical point contribution
\eqref{eq:Gamma-triv-localized}, providing the full Euler--Heisenberg
effective action with the standard principal value prescription at the
poles $T=T_n$.

The \emph{imaginary} part of the effective action comes precisely from the
$\delta$ function term in \eqref{eq:PV-delta-identity}.  Using the coefficient of
the $1/\delta T$ pole obtained in
\eqref{eq:deltaT-pole-integral-final}, we find
\begin{equation}
\operatorname{Im}\Gamma[E]
=
\pi\sum_{n=1}^\infty
\Biggl[
(-1)^{\,n+1}\,
\mathrm{Vol}(\mathbb R^d)\,
\frac{(qE)^{\,\frac d2}}{2(2\pi)^{\,d-1}}\;
\frac{1}{n^{\,\frac d2}}\;
\exp\!\Bigl(-\frac{\pi m^2}{qE}\,n\Bigr)
\Biggr].
\label{eq:Im-Gamma-final}
\end{equation}

\subsubsection{Final result: scalar QED effective action} 

Combining the contribution of the trivial longitudinal critical point
\eqref{eq:Gamma-triv-localized} with the instanton contributions
\eqref{eq:deltaT-pole-integral-final}, the effective action in a constant
electric field $E$ can be written as
\[
\Gamma[E] \;=\; \operatorname{Re}\Gamma[E] \;+\; i\,\operatorname{Im}\Gamma[E].
\]

\paragraph{Real part.}
The real part is given by the localized Euler--Heisenberg proper--time
representation, understood with the principal–value prescription:
\begin{equation}
\operatorname{Re}\Gamma[E]
=
\mathrm{Vol}(\mathbb{R}^{\,d})\,
\mathrm{v.p.}\!\int_0^\infty \frac{dT}{T}\,
(2\pi T)^{-d/2}\,
e^{-\tfrac12 m^2 T}\,
\frac{\tfrac12\,qET}{\sin\!\big(\tfrac12\,qET\big)}.
\label{eq:ReGamma-final}
\end{equation}

\begin{remark}[UV renormalization in $d=4$]
The real part of the effective action \eqref{eq:ReGamma-final} contains
UV divergences from the small–$T$ region of the proper–time integral.
In four dimensions these are removed by the standard local counterterms
\[
\int d^{4}x\,\Bigl(\delta\Lambda\;+\;\delta Z\,F_{\mu\nu}F^{\mu\nu}\Bigr),
\]
corresponding respectively to the cosmological constant and the Maxwell
kinetic term.  Although our derivation uses the Euclidean worldline
representation, it is simply the analytic continuation of the Lorentzian
one–loop effective action, so the renormalization must be imposed with
Lorentz–invariant counterterms.
\end{remark}

\paragraph{Imaginary part (Schwinger pair production).}
The imaginary part induced by the instanton tower at
$T_n = 2\pi n/(qE)$ is
\begin{equation}
\operatorname{Im}\Gamma[E]
=
\mathrm{Vol}(\mathbb{R}^{\,d})\,
\frac{(qE)^{\,\frac d2}}{2(2\pi)^{d-1}}
\sum_{n=1}^{\infty}
\frac{(-1)^{\,n+1}}{n^{\,\frac d2}}\,
\exp\!\Bigl(-\frac{\pi m^2}{qE}\,n\Bigr).
\label{eq:ImGamma-d-final}
\end{equation}
In four spacetime dimensions this becomes
\begin{equation}
\operatorname{Im}\mathcal{L}(E)
=
\frac{(qE)^2}{16\pi^3}
\sum_{n=1}^{\infty}
\frac{(-1)^{\,n+1}}{n^{2}}\,
\exp\!\Bigl(-\frac{\pi m^2}{qE}\,n\Bigr),
\label{eq:ImL-d4-final}
\end{equation}
in exact agreement with Schwinger’s classic result for pair production
in scalar QED.

\begin{remark}[Constant magnetic field]
For a purely magnetic background (e.g.\ $F_{23}=B$, $A_{3}=Bx_{2}$, 
$A_{0}=A_{1}=A_{2}=0$), the Euclidean worldline action 
\eqref{eq:Schwinger-worldline-start} contains the coupling 
$i q A_\mu(x)\dot x^\mu$ with a real gauge potential, and the localizing 
deformation can be applied exactly as in the electric case.  
However, the localization equations admit only the trivial periodic 
solution for all $T>0$; there are no resonant values of $T$ and thus no 
worldline instantons.  
Consequently the effective action is purely real, and localization 
reproduces the standard Euler--Heisenberg expression for a constant 
magnetic field:
\begin{equation}
\Gamma[B]
=
\mathrm{Vol}(\mathbb{R}^{\,d})\,
\int_0^\infty \frac{dT}{T}\,
(2\pi T)^{-d/2}\,
e^{-\tfrac12 m^2 T}\,
\frac{\tfrac12\,qBT}{\sinh\!\big(\tfrac12\,qBT\big)}.
\end{equation}
This reflects the absence of vacuum decay in a static magnetic 
background.
\label{eq:EHB}
\end{remark}

\begin{appendix}

\section{0d LG toy model: Euclidean off-shell SUSY and localization}
\label{sec:0d-LG-toy}

Consider a single real boson \(X\), two Grassmann variables \(\psi,\bar\psi\),
and an auxiliary real field \(F\). Let \(h=h(X)\) and
\(\partial \equiv \frac{d}{dX}\).
We take the Euclidean action
\begin{equation}
\boxed{
S_E[X,\psi,\bar\psi,F]
=\frac{1}{2}\,F^2 \;-\; i\,F\,\partial h(X) \;+\; \partial^2 h(X)\,\bar\psi\,\psi .
}
\label{eq:0d-action}
\end{equation}
Completing the square in \(F\),
\[
\frac12 F^2 - i F\,\partial h
= \frac12\big(F-i\,\partial h\big)^2 + \frac12\,(\partial h)^2 ,
\]
so integrating out the (real) Gaussian \(F\) produces the expected positive
bosonic weight \(\frac12(\partial h)^2\).
Thus \eqref{eq:0d-action} reproduces the familiar on-shell form
\[
S_E \;\longrightarrow\; \frac12\big(\partial h\big)^2
\;+\; \partial^2 h\,\bar\psi\psi .
\]

\paragraph{Off-shell supersymmetry.}
Define a single nilpotent (cohomological) supercharge \(Q\) by
\begin{equation}
\boxed{
Q X=\psi,\qquad Q\psi=0,\qquad Q\bar\psi=i\,F,\qquad QF=0 .
}
\label{eq:0d-Q}
\end{equation}
Then \(Q^2=0\) off shell on all fields. Moreover,
\begin{equation}
\boxed{
S_E \;=\; \frac12\,F^2 \;+\; Q\!\big(\,\bar\psi\,\partial h(X)\,\big).
}
\label{eq:0d-Qexact}
\end{equation}
Indeed,
\[
Q\!\big(\bar\psi\,\partial h\big)
= (Q\bar\psi)\,\partial h + \bar\psi\,\partial^2 h\,QX
= i\,F\,\partial h + \partial^2 h\,\bar\psi\,\psi ,
\]
so \eqref{eq:0d-Qexact} matches \eqref{eq:0d-action}. In particular,
\(Q S_E=0\) (the \(\tfrac12 F^2\) piece is inert since \(QF=0\)).

\paragraph{Localization.}
Consider the \(Q\)-invariant deformation family
\[
Z(t)\;=\;\int dX\,d\psi\,d\bar\psi\,dF\;
\exp\!\Big(-\,\tfrac12 F^2 \;-\; t\,Q(\bar\psi\,\partial h)\Big),
\qquad t>0.
\]
Because \(Q^2=0\) and the measure is \(Q\)-invariant,
\[
\frac{dZ}{dt}
= -\int dX\,d\psi\,d\bar\psi\,dF\;
Q\!\big( \bar\psi\,\partial h\,e^{-\,\frac12 F^2 - t\,Q(\bar\psi\,\partial h)}\big)
= 0,
\]
so \(Z(t)\) is \(t\)-independent.
For large \(t\), the bosonic part is dominated by
\(
\tfrac{t}{2}\,(\partial h)^2
\),
and the integral localizes onto the critical set of \(h\):
\begin{equation}
\boxed{\ \partial h(X)=0\ .\ }
\label{eq:0d-locus}
\end{equation}
Expanding \(X=X_\ast+\xi\) near an isolated nondegenerate critical point
(\(\partial h(X_\ast)=0\), \(\partial^2 h(X_\ast)\neq0\)) gives
\(
\partial h \approx \partial^2 h(X_\ast)\,\xi
\),
so the localized Gaussian yields the standard one-loop factor
\(
\mathrm{sgn}\big(\partial^2 h(X_\ast)\big)
\).
Summing over critical points,
\[
Z \;=\; \sum_{X_\ast:\,\partial h(X_\ast)=0}
\mathrm{sgn}\!\left(\partial^2 h(X_\ast)\right),
\]
the 0d analogue of the Witten index.

\end{appendix}

\bibliographystyle{utphys}
\bibliography{Ref.bib}

\end{document}